\RequirePackage{fix-cm}
\documentclass[a4paper, 12pt, leqno]{article}

\usepackage{latexsym, amsmath, amsthm, amssymb, graphicx, graphics, epsfig}
\usepackage{bm}
\usepackage{tikz-cd}

\tikzset{cong/.style={draw=none,edge node={node [sloped, allow upside down, auto=false]{$\cong$}}},
   Isom/.style={every to/.append style={edge node={node [above,sloped, inner sep=0.4pt, allow upside down, auto=false]{$\sim$}}}}}

\usepackage{fullpage}

\newcommand{\comment}[1]{}

\usepackage[T1]{fontenc}
\usepackage{latexsym}
\usepackage{manfnt}
\usepackage[sans]{dsfont}
\usepackage{palatino}
\usepackage[sc, osf]{mathpazo} 
\usepackage{eulervm}
\usepackage{euscript}

\newcommand{\sub}[1]{\vspace{8pt}\textbf{#1}\hspace*{0.5em}}
\newcommand{\upskip}{\vspace{-8pt}}

\newtheorem{thm}{Theorem}[section]
\newtheorem*{thm*}{Theorem}
\newtheorem{lem}[thm]{Lemma}
\newtheorem*{lem*}{Lemma}

\newtheorem*{cor*}{Corollary}
\newtheorem{prop}[thm]{Proposition}
\newtheorem*{prop*}{Proposition}
\newtheorem*{claim*}{Claim}

\theoremstyle{definition}

\newtheorem{rem}[thm]{Remark}

\newtheorem*{conv*}{Convention}

\numberwithin{equation}{section}

\renewcommand{\proof}{\vspace{-8pt}\noindent\textit{\textbf{Proof. }}}

\renewcommand{\endproof}{$\square$}

\newcommand{\q}[1]{{``#1''}}
\newcommand{\br}[1]{\left(#1\right)}
\newcommand{\brr}[1]{\left[#1\right]}
\newcommand{\bra}[1]{\left\langle #1 \right\rangle}

\renewcommand{\leq}{\leqslant}
\renewcommand{\geq}{\geqslant}

\newcommand{\hookto}{\hookrightarrow}

\newcommand{\set}[1]{\left\{#1\right\}}
\newcommand{\norm}[1]{\left\|#1\right\|}

\newcommand{\inv}{^{-1}}
\newcommand{\restr}[2]{\left. #1 \right|_{#2}}

\newcommand{\matr}[1]{
\begin{pmatrix}
#1
\end{pmatrix}}

\newcommand{\smatr}[1]{\br{ \begin{smallmatrix}#1\end{smallmatrix} } }

\newcommand{\R}{\mathbb R}
\newcommand{\N}{\mathbb N}
\newcommand{\Z}{\mathbb Z}
\newcommand{\CC}{\mathbb C}

\DeclareMathOperator{\Proj}{Proj}
\DeclareMathOperator{\im}{Im}
\DeclareMathOperator{\Ker}{Ker}

\DeclareMathOperator{\Gr}{Gr}

\DeclareMathOperator{\End}{End}
\DeclareMathOperator{\Aut}{Aut}

\DeclareMathOperator{\Id}{Id}

\DeclareMathOperator{\dom}{dom}

\DeclareMathOperator{\Vect}{Vect}
\DeclareMathOperator{\VectM}{Vect^{\infty}_M}
\DeclareMathOperator{\VectpM}{Vect^{\infty}_{\pM}}
\DeclareMathOperator{\VectXM}{Vect_{\,X,\,M}}
\DeclareMathOperator{\VectX}{Vect_{\,X}}
\DeclareMathOperator{\VectXpM}{Vect_{\,X,\,\p M}}
\newcommand{\Ve}{\mathbb V}
\newcommand{\kXM}{k_{X,\,M}}
\newcommand{\kXpM}{k_{X,\,\p M}}

\newcommand{\Reg}{\mathcal R}

\newcommand{\FR}{\mathcal{FR}}
\newcommand{\CRR}{\mathcal{CRR}}
\newcommand{\B}{\mathcal B}
\newcommand{\K}{\mathcal K}

\renewcommand{\H}{\mathcal H}

\newcommand{\F}{\mathcal F}

\newcommand{\E}{\mathcal E}
\newcommand{\A}{\mathcal A}

\renewcommand{\L}{\mathcal L}

\newcommand{\D}{\mathcal D}
\newcommand{\V}{\mathcal V}
\newcommand{\W}{\mathcal W}
\newcommand{\U}{\mathcal U}
\newcommand{\UK}{\U_K}

\newcommand{\Cat}{\mathcal C}

\renewcommand{\mod}{\,\mathrm{mod}\,}

\DeclareMathOperator{\ind}{ind}
\newcommand{\inda}{\ind_{\mathrm{a}}}
\newcommand{\indt}{\ind_{\mathrm{t}}}
\DeclareMathOperator{\Indt}{Ind_{\mathrm{t}}}

\newcommand{\overbar}[1]{\overline{#1\mkern-1mu}\mkern 1mu}

\DeclareMathOperator{\Ell}{Ell}
\newcommand{\Ellt}{\overbar{\Ell}}

\newcommand{\Ellp}{\Ellt^{+}}
\newcommand{\Ellm}{\Ellt^{-}}
\newcommand{\EllXM}{\Ellt_{X,\,M}}
\DeclareMathOperator{\Dir}{Dir}
\DeclareMathOperator{\Dirt}{\overbar{Dir}}

\newcommand{\Dirp}{\overbar{\Dir}^{+}}
\newcommand{\Dirm}{\overbar{\Dir}^{-}}

\newcommand{\Gam}{\Gamma\,}
\newcommand{\Gammabox}{\Gamma^{\boxtimes}\,}
\newcommand{\Gammapm}{\Gamma^{\pm}\,}
\newcommand{\pmbox}{^{\pm\boxtimes}}
\newcommand{\Gammapmbox}{\Gamma\pmbox\;}

\newcommand{\Ebox}{E$^{\boxtimes}$}
\newcommand{\Epm}{E$^{\pm}$}

\newcommand{\Gbox}{G^{\boxtimes}}

\DeclareMathOperator{\spf}{\mathsf{sf}}

\newcommand{\p}{\partial}

\newcommand{\Ginf}{\Gamma^{\infty}}
\newcommand{\Cinf}{C^{\infty}}
\newcommand{\Crinf}{C^{r,\infty}}

\newcommand{\sa}{^{\mathrm{sa}}}

\newcommand{\Em}{E^-}
\newcommand{\Ep}{E^+}

\newcommand{\tM}{\times M}
\newcommand{\pM}{\p M}
\newcommand{\tpM}{\times\pM}
 
\newcommand{\pMS}{\pM\times S^1}
\newcommand{\MX}{X\tM}
\newcommand{\pMX}{X\tpM}

\newcommand{\EY}{E_{\p}}
\newcommand{\EmY}{\Em_{\p}}
\newcommand{\EpY}{\Ep_{\p}}

\newcommand{\Gat}{\widetilde{\Gamma}}

\newcommand{\EEm}{\E^-}

\newcommand{\EEY}{\E_{\p}}
\newcommand{\EEmY}{\EEm_{\p}}

\newcommand{\Tpm}{T$^{\pm}$}
\newcommand{\Tbox}{T$^{\boxtimes}$}

\title{Self-adjoint local boundary problems on compact surfaces. II. Family index} 
\author{Marina\,Prokhorova}
\date{}

\begin{document}

\maketitle

\upskip
\begin{abstract}
The paper presents a first step towards a family index theorem for 
classical self-adjoint boundary value problems.
We address here the simplest non-trivial case of manifolds with boundary, 
namely the case of two-dimensional manifolds.
The first result of the paper is an index theorem for families
of first order self-adjoint elliptic differential operators with local boundary conditions,
parametrized by points of a compact topological space $X$.
We compute the $K^1(X)$-valued index in terms of the topological data over the boundary.
The second result is the universality of the index:
we show that the index is a universal additive homotopy invariant for such families,
if the vanishing on families of invertible operators is assumed.
\end{abstract}

\let\oldnumberline=\numberline
\renewcommand{\numberline}{\vspace{-20pt}\oldnumberline}
\addtocontents{toc}{\protect\renewcommand{\bfseries}{}}

\upskip
\tableofcontents

\section*{Preface}

\addcontentsline{toc}{section}{Preface}
\addtocontents{toc}{\vspace{-18pt}}

An index theory for families of elliptic operators on a closed manifold 
was developed by Atiyah and Singer in \cite{AS71}.
For a family of such operators, parametrized by points of a compact space $X$,
the $K^0(X)$-valued analytical index was computed there in purely topological terms.
An analog of this theory for self-adjoint elliptic operators on closed manifolds 
was developed by Atiyah, Patodi, and Singer in \cite{APS-76};
the index of a family in this case takes values in the $K^1$ group of a base space.

If a manifold has non-empty boundary, the situation becomes more complicated.
The integer-valued index of a single boundary value problem was computed 
by Atiyah and Bott \cite{AB} and Boutet de Monvel \cite{Monvel}.
This result was generalized to the $K^0(X)$-valued index for families of boundary value problems
by Melo, Schrohe, and Schick in \cite{MSS}.

\sub{Manifolds with boundary: self-adjoint case.}
The case of self-adjoint boundary value problems, however, remains largely open.
While Boutet de Monvel's pseudo-differential calculus allows to investigate 
boundary value problems of different types in a uniform manner,
self-adjoint operators seem to lack such a theory.
In this case, two different kinds of boundary conditions, global and local, 
were investigated separately and by different methods.

For Dirac operators on odd-dimensi\-onal manifolds 
with global boundary conditions of Atiya-Patodi-Singer type,
Melrose and Piazza computed the odd Chern character of the $K^1(X)$-valued index \cite{MP2},
which determines the index up to a torsion.
This result is an odd analog of \cite{MP1}.

\sub{Self-adjoint case: local boundary conditions.}
For Dirac operators with classical (that is, local) boundary conditions 
some partial results were obtained in \cite{Pr11, KN, GL, Yu}.

If the base space is a circle, then the analytical index takes values in $K^1(S^1)\cong\Z$ 
and may be identified with the spectral flow.
The spectral flow for curves of Dirac operators over a compact surface 
was computed by the author in \cite{Pr11} in a very special case, 
where all the operators have the same symbol and are considered with the same boundary condition.
Later the results of \cite{Pr11} were generalized to manifolds of higher dimension
by Katsnelson and Nazaikinskii in \cite{KN} and by Gorokhovsky and Lesch in \cite{GL}. 

For compact manifolds of arbitrary dimension, 
Katsnelson and Nazaikinskii expressed the spectral flow 
of a curve of Dirac operators with local boundary conditions
as the (integer-valued) index of the suspension of the curve \cite{KN}.
See also \cite{RS} for a similar result in a more general context.

Gorokhovsky and Lesch considered the straight line between the operators 
$D\otimes\Id$ and $g(D\otimes\Id)g^*$,
where $D$ is a Dirac operator on an even-dimensional compact manifold $M$ 
and $g$ is a smooth map from $M$ to the unitary group $\U(\CC^n)$.
They take the same local boundary condition for all the operators along this line.
Using the heat equation approach, they expressed the spectral flow along this line  
in terms of the spectral flow of boundary Dirac operators \cite{GL}.

The result of Gorokhovsky and Lesch was generalized to the higher spectral flow case by Yu.
In the general situation, the higher spectral flow is defined for a self-adjoint family 
parametrized by points of the product $X=Y\times S^1$ provided that  
the restriction to $Y\times\set{pt}$ has vanishing index in $K^1(Y)$.
The higher spectral flow of such a family takes values in $K^0(Y)$ 
and may be identified with the $K^1(X)$-valued index of the family.
Yu considered a $Y$-parametrized family of Dirac operators on a manifold $M$,
with local boundary conditions, whose $K^1(Y)$-valued index vanishes. 
From such a family and a map $M\to\U(\CC^n)$, 
he constructed a family of straight lines of Dirac operators, as in \cite{GL}.
He expressed the higher spectral flow of such a family 
in terms of the higher spectral flow of the family of boundary Dirac operators \cite{Yu}.

Unfortunately, the methods of \cite{KN, GL, Yu} use essentially the specific nature of Dirac operators
and cannot be applied to more general classes of self-adjoint elliptic differential operators.
In \cite{Pr17} the author generalized results of \cite{Pr11} in a different direction,
computing the spectral flow for curves of \textit{arbitrary} first order self-adjoint elliptic differential operators 
on a compact oriented \textit{surface} with boundary.

\sub{Family index.}
In this paper we generalize the results of \cite{Pr17} to \textit{families} 
of such boundary value problems
parametrized by points of an arbitrary compact space $X$.
Our results may be viewed as a first step towards 
a general family index theorem for classical self-adjoint boundary value problems.

As well as in \cite{Pr17}, we address here the simplest non-trivial case of manifolds with boundary, 
namely the case of \textit{two-dimensional} manifolds.
We consider \textit{first order} self-adjoint elliptic differential operators on such manifolds, 
with \textit{local}, or classical, boundary conditions
(that is, boundary conditions defined by general pseudo-differential operators, 
in particular boundary conditions of Atiyah-Patodi-Singer type, are not allowed).
As it happens, in this setting all the work can be done by topological means only,
without using pseudo-differential calculus.

The first result of the paper is a \textit{family index theorem}:
we define the $K^1(X)$-valued topological index in terms 
of the topological data of the family over the boundary,
and show that the analytical and the topological index coincide.

The second result of the paper is the \textit{universality of the index}
for families of such boundary value problems.
We show that the Grothendieck group of homotopy classes of such families
modulo the subgroup of invertible families is the $K^1$-group of the base space, 
with an isomorphism given by the index.
In fact, we prove stronger results, dealing with the semigroup of such families 
without passing to the Grothendieck group.

\section{Introduction}

\vspace{-5pt}

\sub{Conventions.}
Throughout the paper a \q{Hilbert space} always means a separable complex Hilbert space of infinite dimension,
a \q{compact space} always means a compact Hausdorff topological space,
and a \q{surface} always means a smooth compact oriented connected surface with non-empty boundary.
By the \q{symbol of a differential operator} we always mean its principal symbol.

\sub{Family index for unbounded operators.}
Let $H$ be a Hilbert space.
Denote by $\Reg(H)$ the space of regular (that is, closed and densely defined) 
operators on $H$ equipped with the graph topology.
Recall that this topology (which is also often called the gap topology)
is induced by the metric $\delta\br{A_1, A_2} = \norm{P_1-P_2}$,
where $P_i$ denotes the orthogonal projection of $H\oplus H$ onto the graph of $A_i$. 

Denote by $\Reg\sa(H)$ the subspace of $\Reg(H)$ consisting of self-adjoint operators
and by $\CRR\sa(H)$ the subspace consisting of self-adjoint operators with compact resolvents.

The Cayley transform $A\mapsto \kappa(A) = (A-i)(A+i)\inv$ is a continuous embedding
of $\Reg\sa(H)$ into the unitary group $\U(H)$. 
It takes $\CRR\sa(H)$ into the subgroup $\UK(H)$ of $\U(H)$ 
consisting of unitaries $u$ such that the operator $1-u$ is compact.
Hence $\CRR\sa(H)$ can be considered as a subspace of $\UK(H)$.

As is well known, the group $[X,\UK(H)]$ of homotopy classes 
of maps from a compact topological space $X$ to $\UK(H)$
is naturally isomorphic to $K^1(X)$.
We define the \textit{family index} $\ind(\gamma)$ of a continuous map $\gamma\colon X\to\CRR\sa(H)$
as the homotopy class of the composition $\kappa\circ\gamma\colon X\to\UK(H)$ 
considered as an element of $K^1(X)$,
	\begin{equation*}\label{eq:ind0}
	  \ind(\gamma) = [\kappa\circ\gamma]\in[X,\UK(H)] = K^1(X).
	\end{equation*}
More generally, this definition works as well for graph continuous families of 
regular self-adjoint operators with compact resolvents acting on fibers of a Hilbert bundle over $X$. 
See Section \ref{sec:CRR} for details.

\sub{Local boundary value problems on a surface.}
Throughout the paper $M$ is a fixed smooth compact oriented surface with non-empty boundary $\pM$.
Let $A$ be a first order formally self-adjoint elliptic differential operator
acting on sections of a Hermitian smooth vector bundle $E$ over $M$.
Denote by $\EY$ the restriction of $E$ to $\pM$.
A local boundary condition for $A$ is defined by a smooth subbundle $L$ of $\EY$;
the corresponding unbounded operator $A_L$ on the space $L^2(E)$ of square-integrable sections of $E$ 
has the domain
\begin{equation*}
  \dom\br{A_L} = \set{ u\in H^1(E)\colon \restr{u}{\pM} \mbox{ is a section of } L },
\end{equation*}
where $H^1(E)$ denotes the first order Sobolev space of sections of $E$.
(More precisely, the boundary condition above means that the boundary trace of $u$ is an element of $H^{1/2}(L)$;
see explanation in Section \ref{sec:bound_pr}.)

Let $n$ denote the outward conormal to the boundary $\pM$.
The conormal symbol $\sigma(n)$ of $A$ is self-adjoint 
and thus defines a symplectic structure on the fibers of $\EY$
given by the symplectic 2-form $\omega_x\colon E_x\otimes E_x\to\CC$,
$\omega_x(u,v) = \bra{i\sigma(n_x)\, u, \, v}$ for $x\in \pM$.
Green's formula for $A$ can be written as 
\[
\bra{Au,v}_{L^2(E)} - \bra{u,Av}_{L^2(E)} = \bra{i\sigma(n)\restr{u}{\pM},\restr{v}{\pM}}_{L^2(\EY)}
 = \omega\br{\restr{u}{\pM},\restr{v}{\pM}} 
\text{ for } u,v\in H^1(E).
\]

Let $(n_x,\xi)$ be a positive oriented frame in $T^{*}_x M$, $x\in \pM$.
Since $A$ is elliptic, $\sigma(tn_x+\xi) = t\sigma(n_x) + \sigma(\xi)$ is non-degenerate for every $t\in\R$.
Hence the fiber endomorphism (the symbol of a tangential operator) $b(\xi) = \sigma(n_x)^{-1} \sigma(\xi)$ 
has no eigenvalues on the real axis. 
The generalized eigenspaces $\Ep_x$ and $\Em_x$ of $b(\xi)$ 
corresponding to eigenvalues with positive, resp. negative imaginary part
do not depend on the choice of $\xi$. 
The subspaces $\Ep_x$, resp. $\Em_x$ are fibers of a smooth subbundle $\EpY$, resp. $\EmY$ of $\EY$,
so $\EY$ is naturally decomposed into the direct sum $\EpY\oplus\EmY$.
Moreover, $\EpY$ and $\EmY$ are Lagrangian subbundles of $\EY$
(that is, $\sigma(n)$ takes them to their orthogonal complements).

A local boundary condition $L$ is elliptic for $A$ if 
$L \cap \EpY = L \cap \EmY = 0$ and $L+\EpY = L+\EmY = \EY$;
in this case $A_L$ is a regular operator on $L^2(E)$ with compact resolvents.
If, in addition, $L$ is a Lagrangian subbundle of $\EY$, 
then the regular operator $A_L$ is self-adjoint.
We denote by $\Ellt(E)$ the set of all such pairs $(A,L)$.

\sub{Topological data.}
Following \cite{Pr17},
from an element $(A,L)\in\Ellt(E)$ we extract a topological data which contains 
all the information we need to compute the family index. 
These data are encoded in a vector subbundle $F=F(A,L)$ of $\EY$,
which depends only on $L$ and the restriction of the symbol of $A$ to the boundary.

As was shown by the author in \cite[Proposition 4.3]{Pr17}, 
self-adjoint elliptic local boundary conditions $L$ for $A$
are in a one-to-one correspondence with self-adjoint bundle automorphisms $T$ of $\EmY$.
This correspondence is given by the rule
\begin{equation*}\label{eq:PT0}
	      L=\Ker P_T \; \mbox{ with } \; P_T = P^+\br{ 1+i\sigma(n)\inv TP^- },
\end{equation*}
where $P^+$ denotes the projection of $\EY$ onto $\EpY$ along $\EmY$ and $P^- = 1-P^+$.

Notice that $P_T$ is a bundle projection of $\EY$ onto $\EpY$ along $L$,
so the local boundary condition given by $L$ can be written equivalently in the form $P_T(\restr{u}{\pM}) = 0$
using a bundle projection (a particular case of a pseudo-differential projection), 
as is customary in the study of boundary value problems.

If $A$ is a Dirac type operator, then $\EpY$ and $\EmY$ are mutually orthogonal; 
in this case $L$ can be written (fiber-wise) as
 $L=\set{ u^+\oplus u^-\in \EpY\oplus\EmY \colon \; i\sigma(n)u^+=Tu^- }$.

We associate with a pair $(A,L)\in\Ellt(E)$ the subbundle $F=F(A,L)$ of $\EmY$,
whose fibers $F_x$, $x\in\pM$ are spanned by the generalized eigenspaces of $T_x$ corresponding to negative eigenvalues.

\sub{Analytical index for maps.}
We equip $\Ellt(E)$ with the $C^{1}$-topology on symbols  of operators, the $C^{0}$-topology on their free terms, 
and the $C^{1}$-topology on boundary conditions.
The natural inclusion $\iota\colon\Ellt(E) \hookto  \CRR\sa\br{L^2(E)}$, $(A,L) \mapsto A_L$,
is continuous; see Proposition \ref{prop:AL2}.
For a compact space $X$, this inclusion associates the analytical index 
$\inda(\gamma) := \ind(\iota\circ\gamma)\in K^1(X)$
with every continuous map $\gamma\colon X\to \Ellt(E)$.

\sub{Analytical index for families.} 
More generally, let $\E$ be a locally trivial fiber bundle over $X$,
whose fibers $\E_x$ are smooth Hermitian vector bundles over $M$ 
and the structure group is the group $U(\E_x)$ of smooth unitary bundle automorphisms of $\E_x$
(we equip this group with the $C^1$-topology).
We denote by $\VectXM$ the class of all such bundles $\E$.
Notice that we cannot consider arbitrary vector bundles over $X\times M$,
since we need smoothness with respect to coordinates on $M$.

Let $\Ellt(\E)$ denote the fiber bundle over $X$ associated with $\E$ and having the fiber $\Ellt(\E_x)$ over $x\in X$.
A section of $\Ellt(\E)$ is a family $x\mapsto(A_x, L_x)\in\Ellt(\E_x)$ of operators and boundary conditions 
parametrized by points of $X$.
The natural inclusion $\Ellt(\E_x)\hookrightarrow\CRR\sa(L^2(\E_x))$ 
allows to define the analytical index for such families.
Our first result is computation of the analytical index
in terms of the topological data of a family $(A_x, L_x)$ over $\pM$.

\sub{Topological index.}
With each family $(A_x, L_x)$ as above
we associate its topological index taking values in $K^1(X)$.

Let $\gamma\colon x\mapsto(A_x, L_x)$ be a section of $\Ellt(\E)$.
The family $\br{F(A_x,L_x)}_{x\in X}$ determines the subbundle $\F=\F(\gamma)$ 
of $\EEY = \restr{\E}{X\times\pM}$.
Let $[\F]$ be the class of $\F$ in $K^0(X\times\pM)$.

The second factor $\pM$ is the disjoint union of boundary components $\pM_j$, each of which is a circle. 
Using the natural homomorphism $K^0(X\times S^1) \to K^1(X)$ 
and taking the sum over the boundary components,
we obtain the homomorphism $\Indt\colon K^0(\pMX) \to K^1(X)$.

Finally, we define the topological index of $\gamma$ 
as the value of $\Indt$ evaluated on $[\F]\in K^0(X\times\pM)$, 
$$\indt(\gamma) := \Indt\, [\F(\gamma)] \in K^1(X).$$

\upskip

\sub{Index theorem.}
The first main result of this paper is an index theorem.
It was first announced by the author in \cite{Pr15}.

\textbf{Theorem \ref{thm:ind}.}
\textit{The analytical index of $\gamma$ is equal to its topological index:}
$$\inda(\gamma) = \indt(\gamma).$$
If the base space $X$ is a circle, then $\gamma$ is a one-parameter family of operators.
In this case, up to the identification $K^1(S^1)\cong\Z$,
the analytical index of $\gamma$ coincides with the spectral flow of $\gamma$
and the topological index of $\gamma$ coincides with $c_1(\F(\gamma))[\pMS]$.
Thus for $X=S^1$ our index theorem takes form of \cite[Theorem A]{Pr17}.

\sub{Properties of the analytical index.}
The proof of the index theorem is based on the following properties of the analytical index:
\begin{enumerate}\upskip
	\item[(I0)] Vanishing on families of invertible operators.
	\item[(I1)] Homotopy invariance. 
	\item[(I2)] Additivity with respect to direct sum of operators and boundary conditions.
	\item[(I3)] Functoriality with respect to base changes.
	\item[(I4)] Multiplicativity with respect to twisting by Hermitian vector bundles over the base space.
	\item[(I5)] Normalization: the analytical index of a loop $\gamma\colon S^1 \to\Ellt(E)$ 
	            coincides with the spectral flow of $\gamma$
							up to the natural isomorphism $K^1(S^1)\cong\Z$.
\end{enumerate}\upskip
Here by an \q{invertible operator} we mean a boundary value problem $(A,L)$
such that the unbounded operator $A_L$ has no zero eigenvalues 
(since $A_L$ is self-adjoint, this condition is equivalent to the invertibility of $A_L$).

These properties follow immediately from the analogous properties 
of the family index for unbounded operators on a Hilbert space;
see Section \ref{sec:CRR} for details.
As it happens, these properties alone are sufficient to prove our index theorem.

\sub{Universality of the topological index.}
To describe all invariants of families of self-adjoint elliptic local boundary problems over $M$ 
satisfying properties (I0--I5),
we note first that the topological index satisfies properties (I1--I4).
Property (I0), however, is purely analytical, 
so its connection with the topological index is not clear a priori.
We manage this problem, replacing temporarily (I0) by two topological properties, 
(\Tpm) and (\Tbox), which will be stated below.

First, we replace the subspace $\Ellt^0(E)$ of $\Ellt(E)$ consisting of invertible operators 
by the following two special subspaces of $\Ellt(E)$:
\begin{itemize}\upskip
	\item $\Ellp(E)$ consists of all $(A,L)\in\Ellt(E)$ with positive definite automorphism $T$.
	\item $\Ellm(E)$ consists of all $(A,L)\in\Ellt(E)$ with negative definite automorphism $T$.
\end{itemize}\upskip
Let $\Ellt^0(\E)$, $\Ellp(\E)$, and $\Ellm(\E)$ denote the corresponding subbundles of $\Ellt(\E)$.
We show that every section of $\Ellp(\E)$ or $\Ellm(\E)$ is homotopic to a section of $\Ellt^0(\E)$;
see Propositions \ref{prop:def-pm} and \ref{prop:zero1}.

In addition to this, we consider \q{locally constant} families of operators, 
that is, sections $1_W\boxtimes(A,L)$ of $\Ellt(W\boxtimes E)$, 
where an element $(A,L)\in\Ellt(E)$ is twisted by a vector bundle $W$ over $X$.
See Section \ref{sec:indt-prop} for details.
Since every $(A,L)\in\Ellt(E)$ is connected by a path with an invertible operator,
every section of the form $1_W\boxtimes(A,L)$ is homotopic to a section of $\Ellt^0(W\boxtimes E)$.

Finally, as a substitute for (I0), we take the following two properties:

(\Tpm) Vanishing on sections of $\Ellp(\E)$ and $\Ellm(\E)$.

(\Tbox) Vanishing on \q{locally constant} sections. 

In Section \ref{sec:uni-top} we prove a number of results concerning 
the universal nature of the topological index;
here we show only two of them.

\begin{thm}[this is a particular case of Theorem \ref{thm:top-fam}; see Remark \ref{rem:top-fam}]
\label{thm:top-fam0}
Let $X$ be a compact space and $\Lambda$ be a commutative monoid.
Suppose that we associate an element \mbox{$\Phi(\gamma)\in\Lambda$}
with every section $\gamma$ of $\Ellt(\E)$ for every $\E\in\VectXM$.
Then the following two conditions are equivalent:
\begin{enumerate}\upskip
	\item $\Phi$ satisfies properties (\Tpm, \Tbox) and (I1, I2). 
	\item $\Phi$ has the form $\Phi(\gamma) = \vartheta(\indt(\gamma))$
        for some (unique) monoid homomorphism $\vartheta\colon K^1(X)\to\Lambda$.
\end{enumerate}
\end{thm}

\upskip 

\textbf{Theorem \ref{thm:top-fam2}.}
\textit{
Suppose that we associate an element $\Phi_X(\gamma)\in K^1(X)$
with every section $\gamma$ of $\Ellt(\E)$ 
for every compact space $X$ and every $\E\in\VectXM$.
Then the following two conditions are equivalent:
\begin{enumerate}\upskip
	\item The family $\Phi = (\Phi_X)$ satisfies properties (\Tpm, \Tbox) and (I1--I4).
	\item There is an integer $m$ such that $\Phi = m\cdot\indt$.
\end{enumerate}\upskip
}

\medskip

Theorem \ref{thm:top-fam2} is deduced from Theorem \ref{thm:top-fam0} 
combined with the following result proven in the appendix: 

\textbf{Proposition \ref{prop:K1}.}
\textit{Let $\vartheta$ be a natural self-transformation of the functor $X\mapsto K^1(X)$
	respecting the $K^0(\cdot)$-module structure.	
	Then $\vartheta$ is multiplication by an integer, 
	that is, there is $m\in\Z$ such that
	$\vartheta_X(\mu)=m\mu$ for every $X$ and every $\mu\in K^1(X)$.
	In particular, if $\vartheta_{S^1}$ is the identity, then $\vartheta_X$ is the identity for every $X$.
}

\sub{The proof of the index theorem.}
As was noted above, 
every invariant $\Phi$ satisfying properties (I0) and (I1) satisfies also (\Tpm) and (\Tbox).
Thus Theorem \ref{thm:top-fam2} implies that every invariant $\Phi$ satisfying properties (I0--I4)
has the form $\Phi = m\cdot\indt$.
Applying this to the analytical index, we see that it is an integer multiple of the topological index:
$\inda = m\cdot\indt$ for some integer factor $m=m(M)$, which does not depend on $X$, but can depend on $M$.

To compute $m$, it is sufficient to consider the simplest non-trivial base space, namely $X=S^1$,
where the analytical index is just the spectral flow. 
The spectral flow was computed by the author in \cite{Pr11} for an annulus 
and in \cite{Pr17} for an arbitrary surface.
By \cite[Lemmas 11.3 and 11.5]{Pr17} the value of $m(M)$ is the same for all surfaces $M$,
and by \cite[Theorem 4]{Pr11} $m(M)=1$ for an annulus $M	$. 
These two results together imply that $m(M)=1$ for any surface $M$.
It follows that the analytical index and the topological index of $\gamma$ coincide.

\sub{Universality of the analytical index.} 
The second main goal of the paper is universality of the analytical index.
We obtain a number of results in this direction in Section \ref{sec:uni}, 
combining our index theorem with results of Section \ref{sec:uni-top}.

\sub{Universality for maps.} 
Recall that every complex vector bundle over $M$ is trivial and that $\Ellt(E)$ is empty for bundles $E$ of odd rank.
For $k\in\N$ we denote by $2k_M$ the trivial vector bundle over $M$ of rank $2k$ with the standard Hermitian structure.

\textbf{Theorem \ref{thm:uni-maps-n}.}
\textit{  Let $\gamma\colon X\to\Ellt(2k_M)$, $\gamma'\colon X\to\Ellt(2k'_M)$ be continuous maps.
	Then the following two conditions are equivalent:
\begin{enumerate}\upskip
	\item $\inda(\gamma) = \inda(\gamma')$.
	\item There are $l\in\N$ and maps $\beta\colon X\to\Ellt^0(2(l-k)_M)$, $\beta'\colon X\to\Ellt^0(2(l-k')_M)$ 
	such that $\gamma\oplus\beta$ and $\gamma'\oplus\beta'$ are homotopic as maps from $X$ to $\Ellt(2l_M)$.
\end{enumerate}
}

\upskip
\sub{Semigroup of elliptic operators.}
The disjoint union \[ \Ellt_M = \coprod_{k\in\N}\Ellt(2k_M) \]
has the natural structure of a (non-commutative) graded topological semigroup 
with respect to the direct sum of operators and boundary conditions.
The set $[X,\Ellt_M]$ of homotopy classes of maps from $X$ to $\Ellt_M$ has the induced semigroup structure. 
The semigroup $[X,\Ellt_M]$ is commutative; see Proposition \ref{prop:comm}.

Denote by $\Ellt^0_M = \coprod_{k\in\N}\Ellt^0(2k_M)$ 
the subsemigroup of $\Ellt_M$ consisting of invertible operators.
The inclusion $\Ellt^0_M\hookto\Ellt_M$ induces the homomorphism $[X,\Ellt^0_M]\to[X,\Ellt_M]$;
we will denote by $[X,\Ellt_M]^0$ its image.
The analytical index is homotopy invariant and vanishes on families of invertible operators,
so it factors through $[X,\Ellt_M]/[X,\Ellt_M]^0$.
In other words, there exists a (unique) monoid homomorphism 
$\kappa_a\colon [X,\Ellt_M]/[X,\Ellt_M]^0 \to K^1(X)$ 
such that the following diagram is commutative:
\begin{equation*}
	\begin{tikzcd}
		C(X,\Ellt_M) \arrow{r} \arrow[swap]{drr}{\inda} &
		{[X,\Ellt_M]}  \arrow{r} & 
		{[X,\Ellt_M]/[X,\Ellt_M]^0} \arrow{d}{\kappa_a} 
		\\
		 &  &  K^1(X) 
	\end{tikzcd}
\end{equation*}

\textbf{Theorem \ref{thm:uni-maps2}.}
\textit{The quotient $[X,\Ellt_M]/[X,\Ellt_M]^0$ is an Abelian group isomorphic to $K^1(X)$,
with an isomorphism given by $\kappa_a$.
}

The family index is a universal homotopy invariant for maps from $X$ to $\CRR\sa(H)$,
but the space $\Ellt(E)$ is only a tiny part of $\CRR\sa(L^2(E))$.
Universality is usually lost after passing to a subspace,
so we cannot expect from the analytical index to be a universal invariant for $\Ellt(E)$.
Indeed, it follows from our index theorem that for any given $E$ 
the map $\inda\colon[X,\Ellt(E)]\to K^1(X)$ is neither injective nor surjective for general $X$.
It is surprising that universality can still be restored by considering all vector bundles over $M$ together.

\sub{Universality for families.} 
Denote by $2\kXM\in\VectXM$ the trivial bundle over $X$ with the fiber $2k_M$.

\textbf{Theorem \ref{thm:an-fam}.}
\textit{	Let $\gamma_i$ be a section of $\Ellt(\E_i)$, $i=1,2$.
	Then the following two conditions are equivalent:
\begin{enumerate}\upskip
	\item $\inda(\gamma_1) = \inda(\gamma_2)$.
	\item There are $k\in\N$, sections $\beta^0_i$ of $\Ellt^0(2\kXM)$, 
	and sections $\gamma^0_i$ of $\Ellt^0(\E_i)$ such that 
	$\gamma_1\oplus\gamma^0_2\oplus\beta^0_1$ and 
	$\gamma^0_1\oplus\gamma_2\oplus\beta^0_2$ 
	are homotopic sections of $\E_1\oplus\E_2\oplus 2\kXM$.
\end{enumerate}
}

Let $\Ve$ be a subclass of $\VectXM$ closed under direct sums 
and containing the trivial bundle $2\kXM$ for every $k\in\N$.
In particular, $\Ve$ may coincide with the whole $\VectXM$.

\textbf{Theorem \ref{thm:uni-fam}.}
\textit{Let $X$ be a compact space and $\Lambda$ be a commutative monoid.
Suppose that we associate an element $\Phi(\gamma)\in\Lambda$
with every section $\gamma$ of $\Ellt(\E)$ for every $\E\in\Ve$.
Then the following two conditions are equivalent:
\begin{enumerate}\upskip
	\item $\Phi$ satisfies properties (I0--I2).
	\item $\Phi$ has the form $\Phi(\gamma) = \vartheta(\inda(\gamma))$
        for some (unique) monoid homomorphism\linebreak 
				$\vartheta\colon K^1(X)\to\Lambda$.
\end{enumerate}\upskip
}

\textbf{Theorem \ref{thm:uni-fam2}.}
\textit{Suppose that we associate an element $\Phi_X(\gamma)\in K^1(X)$
with every section $\gamma$ of $\Ellt(\E)$ 
for every compact space $X$ and every $\E\in\VectXM$.
Then the following two conditions are equivalent:
\begin{enumerate}\upskip
	\item The family $\Phi = (\Phi_X)$ satisfies properties (I0--I4).
	\item $\Phi$ has the form $\Phi_X(\gamma) = m\cdot\inda(\gamma)$ for some integer $m$.
\end{enumerate}\upskip
}

Theorems \ref{thm:top-fam2} and \ref{thm:uni-fam2} are stated here as statements about 
the category of compact spaces and continuous maps (maps come into picture due to property (I3)). 
However, these theorems still remain valid if one replaces this category  
by the category of finite CW-complexes and continuous maps
or by the category of smooth closed manifolds and smooth maps. 
The choice of such a category comes into the proofs of these theorems only through
Proposition \ref{prop:K1}, 
and we prove this proposition for each of these three categories.

\sub{Acknowledgments.}
This work is a part of my PhD thesis at the Technion -- Israel Institute of Technology. 
I am grateful to my PhD advisor S.~Reich. 
I am also grateful to N.V.~Ivanov for his support and interest in the work reported in this paper.

\section{Family index for self-adjoint unbounded operators}\label{sec:CRR}

In order to deal with unbounded self-adjoint operators 
(in particular, with self-adjoint differential operators) directly,
one needs an analogue of the  Atiyah--Singer theory \cite{AS}. 
Cf. \cite{BLP-04}, \cite{BJS}, \cite{Jo}.
This section is devoted to such an analogue adapted to our framework.

\sub{The functor $K^1$.}\label{sub:K1}
Let $H$ be a Hilbert space.
Denote by $\B(H)$ the space of bounded linear operators $H\to H$ with the norm topology.

The subspace of unitary operators $\U(H)\subset \B(H)$ is a topological group
with the multiplication defined by composition. 
Let $\UK(H)$ be the subspace  of $\U(H)$ consisting of operators $u$ such that $1-u$ is a compact operator.
It is a closed subgroup of $\U(H)$.

The group structure on $\UK(H)$ induces a (non-commutative) group structure on 
the space $C(X,\UK(H))$ of continuous maps from a compact space $X$ to $\UK(H)$. 
Passing to the set of connected components of $C(X,\UK(H))$
defines a group structure on the set $[X,\UK(H)]$ of homotopy classes of maps from $X$ to $\UK(H)$.
As is well known, the resulting group $[X,\UK(H)]$ is naturally isomorphic to the classical $K^1$-theory
$K^1(X)$ of $X$.
In particular, it is commutative.

\sub{The space of regular operators.}
Recall that an unbounded operator $A$ on $H$ is a linear operator 
defined on a subspace $\D$ of $H$ and taking values in $H$;
the subspace $\D$ is called the domain of $A$ and is denoted by $\dom(A)$.
An unbounded operator $A$ is called closed if its graph is closed in $H\oplus H$ 
and densely defined if its domain is dense in $H$.
It is called \textit{regular} if it is closed and densely defined.

Associating with a regular operator on $H$ the orthogonal projection on its graph
defines an inclusion of the set of regular operators on $H$ into the space 
$\Proj(H\oplus H)\subset\B(H\oplus H)$ of projections in $H\oplus H$.
Let $\Reg(H)$ be the set of regular operators on $H$ together with the topology
induced from
the norm topology on $\Proj(H\oplus H)$ by this inclusion. 
This topology is usually called the \textit{graph topology, or} \textit{gap topology}.
On the subset $\B(H)\subset \Reg(H)$ it coincides with the usual norm topology 
\cite[Addendum, Theorem 1]{CL}.
So, $\B(H)$ is a subspace of $\Reg(H)$; 
it is open and dense in $\Reg(H)$ \cite[Proposition 4.1]{BLP-04}.

A family $\{A_x\}_{x\in X}$ of unbounded operators $A_x\in \Reg(H)$
defined by a family of differential operators and boundary conditions 
with continuously varying coefficients leads to a continuous map $X\to \Reg(H)$.
See, for example, \cite[Appendix A.5]{Pr17}.
This property plays a fundamental role in this circle of questions.

\begin{rem}
Another useful topology on the set of regular operators is the \emph{Riesz topology},
induced by the \emph{bounded transform} $A\mapsto A(1+A^*A)^{-1/2}$ from the norm topology on $\B(H)$.
By definition, the bounded transform takes a Riesz continuous family of regular Fredholm operators 
to a norm continuous family of bounded Fredholm operators, 
so the index of such a family can be defined in a classical way.
The Riesz topology is well suited for the theory of differential operators on closed manifolds,
but, except for several special cases, 
it is unknown whether families of regular operators on $L^2(E)$ 
defined by boundary value problems for sections of $E$ are Riesz continuous.
\end{rem}

\upskip

\sub{Self-adjoint regular operators.}
Recall that the \emph{adjoint operator} of an operator $A\in\Reg(H)$ is an unbounded operator $A^*$ 
with the domain 
$$\dom(A^*) = \set{u\in H\colon \text{ there exists } v\in H \text{ such that } \bra{Aw,u}=\bra{w,v} \text{ for all } w\in H}.$$
For $u\in\dom(A^*)$ such an element $v$ is unique
and $A^*u=v$ by definition.
An operator $A$ is called \emph{self-adjoint} if $A^*=A$
(in particular, $\dom(A^*) = \dom(A)$).

Let $\Reg\sa(H)\subset \Reg(H)$ be the subspace of self-adjoint regular operators.
For $A\in\Reg\sa(H)$, the operator $A+i\colon\dom(A)\to H$ is bijective, 
and the inverse operator $(A+i)\inv$ is bounded \cite[Theorem 3.16]{Kato}.
A self-adjoint regular operator $A$ is said to be an operator \textit{with compact resolvents} 
if $(A+i)\inv$ is a compact operator.
Let $\CRR\sa(H)\subset \Reg\sa(H)$ be the subspace of such 
operators.

\sub{The homotopy type of $\CRR\sa(H)$.}
Booss-Bavnbek, Lesch, and Phillips have shown in \cite{BLP-04}
that the space $\FR\sa(H)$ of Fredholm self-adjoint regular operators is path connected
and that the spectral flow defines the surjective homomorphism 
$$\pi_1(\FR\sa(H))\to\Z.$$
They conjectured that $\FR\sa(H)$ is a classifying space for the functor $K^1$, and
this conjecture was proven by Joachim in \cite{Jo}.
Along the way he proves (crucially using the results of \cite{BJS}) that
$\CRR\sa(H)$ is a classifying space for $K^1$.
In our context $\CRR\sa(H)$ appears to be a more natural choice of classifying space than $\FR\sa(H)$.

The results of Joachim imply that for a compact space $X$ the set of homotopy classes
$[X,\CRR\sa(H)]$ of maps $X\to \CRR\sa(H)$ is naturally isomorphic to $K^1(X)$.
The element of $K^1(X)$ corresponding to a map $\gamma\colon X\to \CRR\sa(H)$
deserves to be called the \emph{family index} of $\gamma$.
At the same time the proofs of the basic properties of this family index
depend on a fairly advanced machinery used in \cite{Jo} and \cite{BJS}, 
and the needed properties are not even stated explicitly in these papers.

By this reason we will use another, more elementary, approach to the family index.
It is based on the Cayley transform and is
a natural development of an idea from \cite{BLP-04}.

\sub{The Cayley transform.}\label{sec:Ca}
The Cayley transform of a self-adjoint regular operator $A$ is the unitary operator defined by the formula 
$$\kappa(A) = (A-i)(A+i)\inv \in\U(H).$$

\begin{prop}\label{prop:Cay1}
  The map $\kappa\colon\Reg\sa(H)\to\U(H)$ is a continuous embedding.
	If $A$ has compact resolvents, then $\kappa(A)\in\UK(H)$. 
\end{prop}

\proof
The first part of the proposition is proven in \cite[Theorem 1.1]{BLP-04}.
The second part follows from the identity $1-\kappa(A) = 2i(A+i)\inv$.
\endproof

\sub{Family index for maps.}
Recall that $K^1(X)=[X,\UK(H)]$.
The Cayley transform 
\begin{equation}\label{eq:CRRUK}
	\kappa\colon\CRR\sa(H)\to\UK(H)
\end{equation}
induces the map 
\begin{equation}\label{eq:CRRK1}
\kappa_*\colon[X,\CRR\sa(H)]\to[X,\UK(H)] = K^1(X).
\end{equation}
It is proved in \cite{Pr19} that \eqref{eq:CRRUK} is a weak homotopy equivalence 
and that the induced map \eqref{eq:CRRK1} is bijective for every compact space $X$.
This motivates our definition of the family index.
Let $\gamma\colon X\to \CRR\sa(H)$ be a continuous map.
We define the \textit{family index} $\ind(\gamma)$ of $\gamma$ as the homotopy class of
the composition $\kappa\circ\gamma\colon X\to\UK(H)$ considered as an element of $K^1(X)$.
In other terms,
\begin{equation}\label{eq:ind}
 \ind(\gamma) = [\kappa\circ\gamma]\in[X,\UK(H)] = K^1(X).
\end{equation}
One can also define in this way the family index of maps $X\to \FR\sa(H)$. 
But for our purposes it is sufficient to consider only maps $X\to \CRR\sa(H)$.

\sub{Families of regular operators.}
More generally, one can consider $X$-parametrized families $(T_x)_{x\in X}$ of regular operators 
acting on a $X$-parametrized family of Hilbert spaces $(\H_x)_{x\in X}$, i.e.
on the fibers of a Hilbert bundle $\H\to X$.

In more details, let $\H\to X$ be a Hilbert bundle, 
that is, a locally trivial fiber bundle over $X$ with a fiber $H$ and the structure group $\U(H)$
(we consider only Hilbert bundles with separable fibers of infinite dimension).
Recall that the
group $\U(H)$ continuously acts on the space $\Reg(H)$ by conjugations:
$(T,g)\mapsto gTg\inv$.
The subspace $\CRR\sa(H)$ is invariant under this action.
This allows to associate with $\H$ the fiber bundle $\CRR\sa(\H)$ having $\CRR\sa(\H_x)$ as the fiber over $x\in X$.
We equip the set $\Gam\CRR\sa(\H)$ of sections of $\CRR\sa(\H)$ with the compact-open topology.

By the Kuiper theorem \cite{Kui}, the unitary group $\U(H)$ is contractible.
Therefore, every Hilbert bundle $\H$ is trivial and a trivialization is unique up to homotopy.
Choice of a trivialization identifies sections of $\CRR\sa(\H)$ with maps from $X$ to $\CRR\sa(H)$.
The \emph{family index} of a section of $\CRR\sa(\H)$ 
is defined as the index of the corresponding map $X\to\CRR\sa(H)$.
This definition does not depend on the choice of trivialization.

\sub{Connection with topological $K$-theory.}
Let again $X$ be a compact space.
The group $K^1(X)$ may also be defined as the direct limit
$\lim_{n\to\infty}[X,\U(\CC^n)]$ with respect to the sequence of embeddings 
\begin{equation}\label{eq:Un}
	\U(\CC^1)\hookto\U(\CC^2)\ldots\hookto\U(\CC^n)\hookto\U(\CC^{n+1})\hookto\ldots
\end{equation}
 given by the rule $u\mapsto u\oplus 1$.

Choice of an orthonormal basis in $H$ allows to identify
\eqref{eq:Un} with a sequence of subgroups of $\U_K(H)$.
By results of Palais \cite{Pal},
the resulting inclusion $j\colon \U_{\infty}\to\UK(H)$ 
of the direct limit $\U_{\infty} = \lim_{n\to\infty}\U(\CC^n)$ is a homotopy equivalence.
In particular, every continuous map $u\colon X\to\UK(H)$ is homotopic 
to a composition $j\circ v$ for some map $v\colon X\to \U_{\infty}$. 
Since $X$ is compact, every map from $X$ to $\U_{\infty}$ takes values in some $\U(\CC^n)$.
Therefore, every map $u\colon X\to\UK(H)$ is homotopic to a map $X\to \U(\CC^n)\subset\UK(H)$ 
for sufficiently large $n$. 
Similarly, if two maps $u,v\colon X\to \U(\CC^n)$ are homotopic as maps to $\UK(H)$,
then they are homotopic as maps to $\U(\CC^m)$ for some $m\geq n$.

The definition of addition in the group $[X,\UK(H)]$ given in the beginning of the section
uses the multiplicative structure of $\UK(H)$.
The standard definition of addition in $\lim_{n}[X,\U(\CC^n)]$
associates with a pair of maps $u,v\colon X\to\U(\CC^n)$ 
the direct sum $u\oplus v \colon X\to\U(\CC^{2n})$,
so that $[u]+[v] = [u\oplus v]\in K^1(X)$.
These two definitions are equivalent, since $u\oplus v$ and $uv\oplus 1$ are homotopic.

Let now $\H$ be a Hilbert bundle over $X$ with a fiber $H$.
The structure group $\U(H)$ of $\H$ acts on $\CRR\sa(H)$ and $\UK(H)$ by conjugations.
The Cayley transform $\kappa\colon\CRR\sa(H)\to\UK(H)$ is equivariant with respect to this action. 
Therefore, $\kappa$ can be applied point-wise to sections of $\CRR\sa(\H)$.
For a section $\gamma$ of $\CRR\sa(\H)$, the Cayley transform $u=\kappa(\gamma)$ is a section of $\UK(\H)$.

Chose a trivialization $J\colon\H\to H_X$,
where $H_X$ denotes the trivial Hilbert bundle $H\times X\to X$.
The composition $u'=J\circ u$ is a map from $X$ to $\UK(H)$ and thus is homotopic to 
$v'\oplus 1$ for some map $v'\colon X\to \U(\CC^n)\subset\UK(H)$.
The classes of $u$ and $v'$ in $K^1(X)$ coincide.
Returning back to $\H$ by applying $J\inv$, we obtain a trivial subbundle $E$ of $\H$ of finite rank 
and a unitary bundle automorphism $v$ of $E$ such that 
the sections $u$ and $v\oplus 1$ of $\UK(\H) = \UK(E\oplus E^{\bot})$ are homotopic.

Conversely, let $E$ be a (not necessarily trivial) vector bundle over $X$.
A bundle automorphism $v$ of $E$ defines an element $[v]\in K^1(X)$ as follows.
Lift $E$ to the product $X\times[0,1]$ and identify the restrictions of $E$ 
to $X\times\set{0}$ and $X\times\set{1}$ twisting the first one by $v$.
This constructions gives a vector bundle over $X\times S^1$ which we denote by $E_v$.
Let $[E_v]$ denote the class of $E_v$ in $K^0(X\times S^1)$.
The group $K^0(X\times S^1)$ is naturally isomorphic to the direct sum $K^0(X)\oplus K^1(X)$;
denote by 
\begin{equation}\label{eq:K0K1}
	\alpha\colon K^0(X\times S^1)\to K^1(X)
\end{equation}
the projection to the second summand.
Then $[v]=\alpha[E_v]\in K^1(X)$.
If $E$ is a subbundle of a Hilbert bundle $\H$ and $u = v\oplus 1$ is a section of $\UK(\H)$,
then $[u]=[v]$ in $K^1(X)$.

\sub{Twisting.} 
One of the key properties of the index that we need in the paper is its multiplicativity 
with respect to twisting by vector bundles.

Let $V$ be a finite-dimensional complex vector space equipped with a Hermitian structure.
An unbounded operator $A$ on $H$ can be twisted by $V$, 
resulting in the unbounded operator $1_V\otimes A$ on $V\otimes H$ 
with the domain $\dom(1_V\otimes A) = V\otimes\dom(A)$.
If an isomorphism $V\cong\CC^d$ is chosen, 
then $V\otimes H$ can be identified with the direct sum of $d$ copies of $H$ 
and $1_V\otimes A$ can be identified with the direct sum of $d$ copies of $A$.
If $A\in\CRR\sa(H)$, then $1_V\otimes A\in\CRR\sa(V\otimes H)$.

Let now $W$ be a finite rank Hermitian vector bundle over $X$.
A Hilbert bundle $\H$ over $X$ can be twisted by $W$, 
giving rise to another Hilbert bundle $W\otimes\H$ over $X$ 
with the fiber $(W\otimes\H)_x = W_x\otimes\H_x$ over $x\in X$.
A section $\gamma$ of $\CRR\sa(\H)$ can be twisted by $W$, 
resulting in the section $1_W\otimes\gamma$ of $\CRR\sa(W\otimes\H)$
such that $(1_W\otimes\gamma)(x) = 1_{W_x}\otimes\gamma(x)$.
Since the Cayley transform is additive with respect to direct sums 
and equivariant with respect to conjugation by unitaries,
$\kappa(1_W\otimes\gamma) = 1_W\otimes\kappa(\gamma)$. 

Chose a subbundle $E\subset\H$ of finite rank and a unitary bundle automorphism $v$ of $E$ such that 
the sections $\kappa(\gamma)$ and $v\oplus 1$ of $\UK(\H)$ are homotopic.
Then the sections $1_W\otimes \kappa(\gamma)$ and $(1_W\otimes v)\oplus 1_{W\otimes E^{\bot}}$ 
of $\U_K(W\otimes\H)$ are also homotopic.
The vector bundle $(W\otimes E)_{1_W\otimes v}$ is isomorphic to $p^*W\otimes E_v$, 
where $p$ denotes the projection $X\times S^1\to X$.
Since \eqref{eq:K0K1} is a homomorphism of $K^0(X)$-modules, we get 
\[
   [1_W\otimes v] = \alpha[(W\otimes E)_{1_W\otimes v}] 
	= \alpha(p^*[W]\cdot[E_v]) = [W]\cdot\alpha[E_v] = [W]\cdot[v]\in K^1(X).
\]
It follows that
$$\ind(1_W\otimes \gamma) = [1_W\otimes \kappa(\gamma)] = [1_W\otimes v]
 = [W]\cdot[v] = [W]\cdot[\kappa(\gamma)] = [W]\cdot\ind(\gamma)\in K^1(X).$$

\sub{Properties of the family index.}
In fact, we do not need an exact definition of the family index to prove the main results of the paper.
All we need is the following properties of the index.

\begin{prop}\label{prop:I0I5}
\label{prop:prop-inda}
The family index has the following properties for every compact spaces $X$, $Y$ 
and Hilbert bundles $\H$, $\H_0$, $\H_1$ over $X$.
\begin{enumerate}\upskip
	\item[(I0)] Vanishing. The index of a family of invertible operators vanishes.
	\item[(I1)] Homotopy invariance. If $\gamma$ and $\gamma'$ are homotopic sections of $\CRR\sa(\H)$, 
   	then $\ind(\gamma) = \ind(\gamma')$.
	\item[(I2)] Additivity.
	            $\ind(\gamma_0\oplus\gamma_1) = \ind(\gamma_0)+\ind(\gamma_1)$ 
							for every sections $\gamma_i$ of $\CRR\sa(\H_i)$, $i=0,1$.
	\item[(I3)] Functoriality. Let $f\colon Y\to X$ be a continuous map and $\gamma$ be a section of $\CRR\sa(\H)$. 
	            Then $\ind(f^*\gamma) = f^*\ind(\gamma) \in K^1(Y)$,
							where $f^*\gamma = \gamma\circ f$ is the section of $\CRR\sa(f^*\H)$.
	\item[(I4)] Twisting. $\ind(1_W\otimes\gamma) = [W]\cdot\ind(\gamma)$ for every section $\gamma$ of $\CRR\sa(\H)$
	            and every Hermitian vector bundle $W$ over $X$, 
							where $[W]$ denotes the class of $W$ in $K^0(X)$.
	\item[(I5)] Normalization. For a loop $\gamma\colon S^1 \to\CRR\sa(H)$, 
	the index of $\gamma$ coincides with the spectral flow of $\gamma$
	up to the natural isomorphism $K^1(S^1)\cong\Z$.
	\item[(I6)] Conjugacy invariance.
	            The index of a section of $\CRR\sa(\H)$ is invariant with respect to the conjugation  
							by a unitary bundle automorphism of $\H$.
							In other words, $\ind (u\gamma u^*) = \ind(\gamma)$ for every 
							section $\gamma$ of $\CRR\sa(\H)$ and every section $u$ of $\U(\H)$.
\end{enumerate}
\end{prop}

\proof
(I1) and (I3) follows immediately from the definition of the index.
(I4) is proven in the previous subsection.

(I0). The Cayley transform takes the subspace of $\CRR\sa(H)$ consisting of invertible operators 
to the subspace $\UK^0(H) = \set{u\in\UK(H)\colon u+1 \text{ is invertible}}$ of $\UK(H)$.
The space $\UK^0(H)$ is contractible, with the contraction given by the formula $h_t(u) = \exp(t\log(u))$,
where $\log\colon \U(\CC)\setminus\set{-1}\to i(-\pi,\pi)\subset i\R$ is the branch of the natural logarithm.
Therefore, for every section $\gamma$ of $\CRR\sa(\H)$ consisting of invertible operators 
the composition $\kappa\circ\gamma$ is a section of $\UK^0(\H)$ homotopic to the identity section,
so $\ind(\gamma) = [\kappa\circ\gamma] = 0$.

(I2). 
Let $u_i = \kappa(\gamma_i)$.
The Cayley transform is additive with respect to direct sums, 
so $\kappa(\gamma_0\oplus\gamma_1) = \kappa(\gamma_0)\oplus \kappa(\gamma_1)$. 
Let $E_i$ be a trivial subbundle of $\H_i$ of finite rank 
and $v_i$ be a unitary bundle automorphism of $E_i$ such that 
the sections $\kappa(\gamma_i)$ and $v_i\oplus 1$ of $\UK(\H_i)$ are homotopic.
Then $\kappa(\gamma_0)\oplus \kappa(\gamma_1)$ and $(v_0\oplus v_1)\oplus 1$ are also homotopic,
and $\ind(\gamma_0\oplus\gamma_1) = [v_0\oplus v_1] = [v_0]+[v_1] = \ind(\gamma_0)+\ind(\gamma_1)$.

(I5) follows from \cite[Proposition 2.17]{BLP-04}.

(I6). Since the unitary group of a Hilbert space is contractible, 
there is a homotopy $(u_t)_{t\in[0,1]}$ connecting $u_0=1$ and $u_1=u$.
It induces the homotopy $v_t = u_t v u_t^*$ connecting the sections 
$v=\kappa(\gamma)$ and $uvu^*$ of $\UK(\H)$.
Therefore, $\ind(u\gamma u^*) = [uvu^*] = [v] = \ind(\gamma)\in K^1(X)$.
\endproof

\section{Elliptic local boundary value problems}\label{sec:bound_pr}

Throughout the paper $M$ is a smooth compact connected oriented surface 
with non-empty boundary $\pM$ and a fixed Riemannian metric.

\sub{Operators.}
Denote by $\Ell(E)$ the set of first order formally self-adjoint elliptic differential operators acting on sections
of a smooth Hermitian complex vector bundle $E$ over $M$.
Recall that an operator $A$ is called elliptic if its (principal) symbol
$\sigma_A(\xi)$ is non-degenerate for every non-zero cotangent vector $\xi\in T^{\ast}M$.
An operator $A$ is called formally self-adjoint if it is symmetric on the domain $C^{\infty}_0(E)$,
that is, if $\int_M\bra{Au,v}ds = \int_M\bra{u,Av}ds$
for any smooth sections $u$, $v$ of $E$ with compact supports in $M\setminus\pM$.
Throughout the paper all differential operators are supposed 
to have smooth ($C^{\infty}$) coefficients.

\sub{Local boundary conditions.}
The differential operator $A\in\Ell(E)$ with the domain $C^{\infty}_0(E)$
is a symmetric unbounded operator on the Hilbert space $L^2(E)$ of $L^2$-sections of $E$.
This operator can be extended to a regular self-adjoint operator
on $L^2(E)$ by imposing appropriate boundary conditions.
We will consider only local boundary conditions.
Denote by $\EY$ the restriction of $E$ to the boundary $\pM$ of $M$.
A smooth subbundle $L$ of $\EY$ defines a local boundary condition for $A$;
the corresponding unbounded operator $A_L$ on $L^2(E)$ has the domain
\begin{equation}\label{eq:dom}
  \dom\br{A_L} = \set{ u\in H^1(E)\colon \restr{u}{\pM} \mbox{ is a section of } L },	
\end{equation}
where $H^1(E)$ denotes the first order Sobolev space
(the space of sections of $E$ which are in $L^2$ together with all their first derivatives).
We will often identify a pair $(A,L)$ with the operator $A_L$.

To give a precise meaning to the notation in the right-hand side of \eqref{eq:dom},
recall that the restriction map $\Cinf(E)\to \Cinf(\EY)$ 
taking a section $u$ to $\restr{u}{\pM}$
extends continuously to the trace map $\tau\colon H^1(E)\to H^{1/2}(\EY)$.
The smooth embedding $L\hookto\EY$ defines the natural inclusion $H^{1/2}(L)\hookto H^{1/2}(\EY)$.
By the condition \q{$\restr{u}{\pM}$ is a section of $L$} in \eqref{eq:dom}
we mean that the trace $\tau(u)$ lies in the image of this inclusion.

\sub{Decomposition of $E$.}
To describe when a subbundle $L$ is an \q{appropriate boundary condition},
give first some properties of self-adjoint elliptic symbols on a surface.

\begin{prop}[\cite{Pr17}, Proposition 4.1]\label{prop:Q}
Let $\sigma$ be the symbol of an operator $A\in\Ell(E)$.
Then the rank of $E$ is even and $E$ is naturally decomposed into the direct (not necessarily orthogonal) 
sum $E = \Ep \oplus \Em$ of two smooth subbundles $\Ep=\Ep(\sigma)$ and $\Em=\Em(\sigma)$ 
of equal rank satisfying the following conditions.
For any positive oriented frame $(e_1, e_2)$ in $T^{*}_x M$, $x \in M$,
the fibers $\Ep_x$ and $\Em_x$ are invariant subspaces of the operator $Q_x = \sigma(e_1)^{-1} \sigma(e_2)\in \End(E_x)$. 
All eigenvalues of the restriction of $Q_x$ to $\Ep_x$, resp. $\Em_x$ have positive, resp. negative imaginary part.
Finally, $\sigma(\xi)\Ep_x = (\Ep_x)^{\bot}$ and $\sigma(\xi)\Em_x = (\Em_x)^{\bot}$ for any non-zero $\xi\in T^{*}_x M$.
\end{prop}

\upskip

\sub{Self-adjoint elliptic boundary conditions.}
Denote $\EmY = \restr{\Em}{\pM}$ and $\EpY = \restr{\Ep}{\pM}$.
Let $n$ be the outward conormal to $\pM$.
The conormal symbol $\sigma(n)$ of $A$ defines a symplectic structure on the fibers of $\EY$
given by the symplectic 2-form
$\omega_x(u,v) = \bra{i\sigma(n) u,v}$ for $u,v\in E_x$, $x\in \pM$.
With respect to this symplectic structure,
$\EpY$ and $\EmY$ are Lagrangian subbundles of $\EY$.

A smooth subbundle $L$ of $\EY$  is an \textbf{elliptic boundary condition} for $A$
(or, what is one and the same, Shapiro-Lopatinskii boundary condition) if
\begin{equation}\label{eq:L1}
L \cap \EpY = L \cap \EmY = 0 \;\text{ and }\; L+\EpY = L+\EmY = \EY.
\end{equation}
If additionally $L$ is a Lagrangian subbundle of $\EY$, 
that is, $\sigma(n) L = L^{\bot}$,
then $L$ is a \textbf{self-adjoint boundary condition} for $A$.
See e.g. \cite[Sections 3 and 4]{Pr17} for more detailed description.

We denote by $\Ellt(E)$ the set of all pairs $(A,L)$ such that $A\in\Ell(E)$
and $L$ is a smooth Lagrangian subbundle of $\EY$ satisfying condition \eqref{eq:L1}. 

\begin{prop}[\cite{Pr17}, Proposition 4.2]
\label{prop:AL1}
For every $(A,L)\in\Ellt(E)$ the unbounded operator $A_L$ 
is a regular self-adjoint operator on $L^2(E)$ with compact resolvents.
\end{prop}

\section{The analytical index}\label{sec:Ellt}

For a smooth complex vector bundle $V$ over a smooth manifold $N$, 
we denote by $\Gr(V)$ the smooth bundle over $N$
whose fiber over $x\in N$ is the complex Grassmanian $\Gr(V_x)$.
In the same manner we define the smooth bundle $\End(V)$ of fiber endomorphisms. 
We will identify sections of $\Gr(V)$ with subbundles of $V$ 
and sections of $\End(V)$ with bundle endomorphisms of $V$.

\sub{The topology on $\Ellt(E)$.}
We equip $\Ell(E)$ with the $C^{1}$-topology on symbols and the $C^{0}$-topology on free terms of operators.
To be more precise, notice that 
$M$ is homotopy equivalent to a wedge of circles, so the tangent bundle $TM$ is trivial.
Hence we can choose smooth global sections $e_1$, $e_2$ of $TM$ such that
$(e_1(x), e_2(x))$ is an orthonormal frame of $T_x M$ for any $x\in M$.
Choose a smooth unitary connection $\nabla$ on $E$.
Each $A\in\Ell(E)$ can be written uniquely as 
$A = a_1\nabla_1 + a_2\nabla_2 + a$,	where $\nabla_i = \nabla_{e_i}$ 
and $a_1$, $a_2$, $a$ are bundle endomorphisms.
Therefore the choice of $(e_1,e_2,\nabla)$ defines the inclusion
$$\Ell(E) \hookto \Cinf\br{\End(E)}^2 \times \Cinf\br{\End(E)},
  \quad a_1\nabla_1 + a_2\nabla_2 + a \mapsto \br{(a_1, a_2), a},$$
	where $\Cinf\br{\End(E)}$ denotes the space of smooth sections of $\End(E)$.
We equip $\Ell(E)$ with the topology induced by the inclusion 
$$\Ell(E) \hookto C^{1}\br{\End(E)}^2 \times C^{0}\br{\End(E)}$$
and equip $\Ellt(E)$ with the topology induced by the inclusion
$$\Ellt(E) \hookto \Ell(E)\times C^{1}(\Gr(\EY))$$
(with the product topology on the last space).
Thus defined topologies on $\Ell(E)$ and $\Ellt(E)$ are independent of the choice 
of a frame $(e_1,e_2)$ and a connection $\nabla$.

\begin{prop}[\cite{Pr17}, Proposition 5.1]\label{prop:AL2}
The natural inclusion 
\begin{equation}\label{eq:iota}
  \iota\colon\Ellt(E) \hookto  \CRR\sa\br{L^2(E)}, \quad (A,L) \mapsto A_L,	
\end{equation}
is continuous.
\end{prop}

\upskip
\sub{The analytical index of a map.}
Let $\gamma$ be a continuous map from a compact topological space $X$ to $\Ellt(E)$.
We define \textit{the analytical index of} $\gamma$ to be the index 
of the composition of $\gamma$ with the inclusion $\iota\colon \Ellt(E) \hookto \CRR\sa(L^2(E))$
and will denote it by $\inda(\gamma)$.

More generally, the index can be defined for a family of elliptic operators acting on a family of bundles; 
we describe such a situation below.

\sub{Families of elliptic operators.}
For a smooth Hermitian vector bundle $E$ over $M$,
we denote by $U(E)$ the group of smooth unitary bundle automorphisms of $E$ with the $C^1$-topology.

The continuous action of the topological group $U(E)$ on $E$ induces 
the continuous embedding $U(E) \hookto \U(L^2(E))$.
The action of $U(E)$ on $\Ellt(E)$ given by the rule $g(A,L) = (gAg\inv, \; gL)$
is continuous and compatible with the action of $U(E)$ on $\Reg(L^2(E))$.

Denote by $\VectX$ the class of all Hermitian vector bundles over $X$ 
and by $\VectM$ the class of all smooth Hermitian vector bundles over $M$.
Denote by $\VectXM$ the class of all locally trivial fiber bundles over $X$
with fibers $\E_x\in\VectM$ and the structure group $U(\E_x)$.
Note that in the case of disconnected $X$ the fibers over different points of $X$ 
are not necessarily isomorphic.

Let $\E\in\VectXM$. 
We will denote by $\Ellt(\E)$ the locally trivial fiber bundle over $X$ 
with the fiber $\Ellt(\E_x)$ associated with $\E$.
A section of $\Ellt(\E)$ is just a family of elliptic operators 
acting on fibers of a family $(\E_x)$ of vector bundles over $M$ parametrized by points of $X$.
We denote by $\Gam\Ellt(\E)$ the space of sections of $\Ellt(\E)$ equipped with the compact-open topology.

\sub{The analytical index of a family.} 
A bundle $\E\in\VectXM$ defines the Hilbert bundle $\H=\H(\E)$ over $X$, 
whose fiber over $x\in X$ is $\H_x=L^2(\E_x)$.
Note that the fibers $\H_x$ over different points $x$ are isomorphic as Hilbert spaces
even if $\E_x$ are not isomorphic as vector bundles over $M$.

The natural embedding $\iota\colon \Ellt(E) \hookto \CRR\sa(L^2(E))$ is $U(E)$-equivariant 
and thus defines the bundle embedding $\Ellt(\E) \hookto \CRR\sa(\H)$, 
which we still will denote by $\iota$.
For a section $\gamma$ of $\Ellt(\E)$, $\iota(\gamma)$ is a section of $\CRR\sa(\H)$.
The \textit{analytical index} $\inda(\gamma)$ of $\gamma$ is defined as the family index of $\iota(\gamma)$.

\upskip
\sub{Invertible operators.}
We denote by $\Ellt^0(E)$ the subspace of $\Ellt(E)$ consisting of all pairs $(A,L)$
such that the unbounded operator $A_L$ has no zero eigenvalues 
(since $A_L$ is self-adjoint, this condition is equivalent to the invertibility of $A_L$).
For $\E\in\VectXM$ we denote by $\Ellt^0(\E)$ the subbundle of $\Ellt(\E)$, whose fiber over $x\in X$ is $\Ellt^0(\E_x)$.

Property (I0) of Proposition \ref{prop:prop-inda} implies 
that the analytical index vanishes on sections of $\Ellt^0(\E)$;
our proof of the index theorem will rely heavily upon this fact.

\section{The topological index}\label{sec:indt}

The first main result of the paper is computation of the analytical index 
of a section $\gamma\colon x\mapsto(A_x,L_x)$ of $\Ellt(\E)$
in terms of topological data of $\gamma$ over the boundary.
These data are encoded in the family $\F = (\F_x)_{x\in X}$ of vector bundles over $\pM$ with $\F_x = F(A_x,L_x)$.

\sub{The correspondence between boundary conditions and automorphisms of $\EmY$.}
Let $A\in\Ell(E)$.
Define $\Ep=\Ep(A)$ and $\Em=\Em(A)$ as in Proposition \ref{prop:Q}.
Let $\EpY$, resp. $\EmY$ be the restriction of $\Ep$, resp. $\Em$ to the boundary $\pM$.

Suppose for a moment that $\EpY$ and $\EmY$ are \textit{mutually orthogonal} subbundles of $\EY$
(this holds, in particular, for Dirac type operators).
With every subbundle $L\subset\EY$ satisfying condition \eqref{eq:L1}
we can associate the projection of  $\EmY$ onto $\EpY$ along $L$.
Composing this projection with $-i\sigma(n)\colon\EpY\to(\EpY)^{\bot}=\EmY$, we obtain the bundle automorphism $T$ of $\EmY$.
Conversely, with every bundle automorphism $T$ of $\EmY$ we associate the subbundle $L$ of $\EY$ 
given by the formula 
\begin{equation}\label{eq:ortL}
		L=\set{ u^+\oplus u^-\in \EpY\oplus\EmY = \EY \colon \; i\sigma(n)u^+=Tu^- }.
\end{equation}
The automorphism $T$ is self-adjoint if and only if $L$ is Lagrangian, 
so we obtain a bijection between the set of all self-adjoint elliptic local boundary conditions for $A$
and the set of all self-adjoint bundle automorphisms of $\EmY$.

This simple trick does not work in the general case, where $\EpY$ and $\EmY$ are not mutually orthogonal.
However, it can be modified to obtain such a bijection for the general case as well,
though in a bit more complicated manner.
Namely, we associate with $L$ an automorphisms $T$ of $\EmY$ making the following diagram commutative. 
See \cite[Section 4]{Pr17} for details.
\begin{equation*}\label{diag:T}							
\begin{tikzcd}
L \arrow{d}[swap]{P^-} \arrow{r}{P^+} & \EpY \arrow{r}{i\sigma(n)} 
& (\EpY)^{\bot} \arrow[swap]{d}{P^-_{\mathrm{ort}}} \\
\EmY \arrow[swap,dashed]{rr}{T} 
&& \EmY \arrow[swap, shift right=1ex]{u}{(P^-)^{\ast}} 
\end{tikzcd}
\end{equation*}

\begin{prop}[\cite{Pr17}, Proposition 4.3]\label{prop:T}
Let $A\in\Ell(E)$.
Denote by $P^+$ the projection of $\EY$ onto $\EpY$ along $\EmY$
and by $P^- = 1-P^+$ the projection of $\EY$ onto $\EmY$ along $\EpY$.
Then the following statements hold.
\begin{enumerate}\upskip
	\item There is a one-to-one correspondence between 
	      smooth subbundles $L$ of $\EY$ satisfying condition \eqref{eq:L1}
				and smooth bundle automorphisms $T$ of $\EmY$.
				This correspondence is given by the formula
	      \begin{equation}\label{eq:PT}
	      L=\Ker P_T \; \mbox{ with } \; P_T = P^+\br{ 1+i\sigma(n)\inv TP^- };
	      \end{equation}
	      here $P_T$ is the projection of $\EY$ onto $\EpY$ along $L$.
	\item For $L$ and $T$ as above, $L$ is Lagrangian if and only if $T$ is self-adjoint.
\end{enumerate}\upskip
If $\EpY$ and $\EmY$ are mutually orthogonal, then \eqref{eq:PT} is equivalent to \eqref{eq:ortL}.
\end{prop}

\noindent
It is shown in \cite[Proposition 5.3]{Pr17} that the correspondence $(A,L)\mapsto(A,T)$ is a homeomorphism.
This allows us to move freely from $(A,L)$ to $(A,T)$ and back; 
we will use it further without special mention in constructions of homotopies.

\sub{Definition of $F(A,L)$.}
The map $F$ from $\Ellt(E)$ to the space of smooth subbundles of $\EY$ is defined as follows.
Let $(A,L) \in \Ellt(E)$ and $T$ be the self-adjoint automorphism of $\EmY$ given by formula \eqref{eq:PT}.
We define $F_x$ as the invariant subspace of $T_x$
spanned by the generalized eigenspaces of $T_x$ corresponding to negative eigenvalues.
Subspaces $F_x$ of $\Em_x$ smoothly depend on $x\in\pM$
and therefore are fibers of a smooth subbundle $F = F(A,L)$ of $\EmY$.
Being a subbundle of $\EmY$, $F(A,L)$ is also a smooth subbundle of $\EY$.
Moreover, the map $F \colon \Ellt(E) \to C^1(\Gr(\EY))$ is continuous; 
see \cite[Proposition 5.3]{Pr17}.

\sub{Subbundles, restrictions, and forgetting of smooth structure.}
For $\V\in\Vect_{\,X,\,M}$ we denote by $\V_{\p}\in\VectXpM$ 
the locally trivial bundle over $X$ whose fiber over $x\in X$ is the restriction of $\V_x$ to $\pM$.

Let $N$ be a smooth manifold (in our case it will be either $M$ or $\pM$),
and let $\V\in\Vect_{\,X,\,N}$.
We say that $\W\subset \V$ is a \textit{subbundle} of $\V$
if $\W\in\Vect_{\,X,\,N}$ and $\W_x$ is a smooth subbundle of $\V_x$ for every $x\in X$.

We will denote by $\bra{\V}$ the vector bundle over $X\times N$
whose restriction to $\set{x}\times N$ is the fiber $\V_x$ with forgotten smooth structure.

\sub{Definition of $\F(\gamma)$.} 
Let $\gamma$ be a section of $\Ellt(\E)$, $\E\in\VectXM$.
By \cite[Propositions 5.2 and 5.3]{Pr17}, $\Em(A_x)$ and $F(A_x,L_x)\subset\Em(A_x)$ continuously depend on $x$.
Hence they define the subbundle $\EEm(\gamma)$ of $\E$ whose fiber over $x$ is $\Em(A_x)$
and the subbundle $\F=\F(\gamma)$ of $\EEmY(\gamma)$ whose fiber over $x$ is $F(A_x,L_x)$.

\sub{The homomorphism $\Indt$.}
The boundary $\pM$ is a disjoint union of circles, 
so the natural homomorphism 
$$K^0(X)\otimes K^0(\pM) \oplus K^1(X)\otimes K^1(\pM) \longrightarrow K^0(\pMX)$$
is an isomorphism.
Denote by $\alpha_{\p}$ the projection of $K^0(\pMX)$ on the second summand $K^1(X)\otimes K^1(\pM)$ of this direct sum.
The orientation of $M$ induces an orientation of $\pM$ and thus defines the identification 
of $K^1(\pM) = \bigoplus_{j=1}^m K^1(\pM_j)$ with $\Z^m$,
where $\pM_j$, $j=1\ldots m$, are the boundary components. 
Denote by $\delta$ the homomorphism $K^1(\pM) = \Z^m\to\Z$ given by the formula 
$(a_1,\ldots,a_m)\mapsto\sum_{j=1}^m a_j$.
Equivalently, $\delta$ is the connecting homomorphism of the exact sequence 
\begin{equation*}\label{eq:delta}
  \begin{tikzcd}
	  K^1(M) \arrow{r}{i^{\ast}} & K^1(\pM) \arrow{r}{\delta} & K^0(M,\pM) = \Z,
	\end{tikzcd}
\end{equation*}
where $i$ denotes the inclusion $\pM\hookrightarrow M$
and the identification of $K^0(M,\pM)$ with $\Z$ is given by the orientation of $M$.

We define the topological index homomorphism 
$$\Indt\colon K^0(\pMX)\to K^1(X)$$ 
to be the composition 
\begin{equation}\label{eq:beta}
  \begin{tikzcd}
	  K^0(\pMX) \arrow{r}{\alpha_{\p}} & K^1(X)\otimes K^1(\pM) 
  \arrow{r}{\Id\otimes\delta} & K^1(X) \otimes\Z  = K^1(X).
	\end{tikzcd}  
\end{equation}

\sub{The topological index.}
We define the topological index of a section $\gamma$ of $\Ellt(\E)$ by the formula 
\begin{equation}\label{eq:indt}
  \indt(\gamma) = \Indt\,[\F(\gamma)],
\end{equation}
where $[\F]$ denotes the class of $\bra{\F}$ in $K^0(\pMX)$.

\section{Properties of the topological index}\label{sec:indt-prop}

\upskip

\sub{Properties of the homomorphism $\Indt$.}
Denote by $G^{\p}$ the image of the homomorphism $K^0(\MX) \to K^0(\pMX)$
induced by the embedding of $\pMX$ to $\MX$.

Denote by $G^{\boxtimes}$ the image of the natural homomorphism $K^0(X)\otimes K^0(\pM) \to K^0(\pMX)$.
Recall that this homomorphism takes the tensor product $[W]\otimes[V]$ 
of the classes of vector bundles $W$ over $X$ and $V$ over $\pM$
to the class of their external tensor product $[W\boxtimes V]\in K^0(\pMX)$.

Denote by $G$ the subgroup of $K^0(\pMX)$ spanned by $G^{\p}$ and $G^{\boxtimes}$.

\begin{prop}\label{prop:Indt}
The homomorphism $\Indt$ is surjective with the kernel $G$. 
In other words, the following sequence is exact:
\begin{equation*}
	\begin{tikzcd}
	  0 \arrow{r} & G \arrow{r} & K^0(\pMX) \arrow{r}{\Indt} & K^1(X) \arrow{r} & 0.
	\end{tikzcd}
\end{equation*}
\end{prop}

\proof
The groups $K^{\ast}(M)$ and $K^{\ast}(\pM)$ are free of torsion, 
so the first two rows of the following commutative diagram are short exact sequences:
\begin{equation*}
	\begin{tikzcd}
			0 \arrow{r} & K^0(X)\otimes K^0(M) \arrow{r}         \arrow{d}{\Id\otimes i^{\ast}} & 
			              K^0(\MX)             \arrow{r}{\alpha} \arrow{d}{(\Id\times i)^{\ast}} & 
										K^1(X)\otimes K^1(M) \arrow{r} \arrow{d}{\Id\otimes i^{\ast}} & 0 
			\\
			0 \arrow{r} & K^0(X)\otimes K^0(\pM) \arrow{r} & 
			              K^0(\pMX) \arrow{r}{\alpha_{\p}} \arrow{rd}[swap]{\Indt} &
			              K^1(X)\otimes K^1(\pM) \arrow{d} \arrow{d}{\Id\otimes\delta} \arrow{r} & 0
			\\
			\; & \; & \; & K^1(X)\otimes\Z \arrow{d} & 
			\\
			\; & \; & \; & 0 & 
	\end{tikzcd}
\end{equation*}
\upskip
Taking tensor product of the exact sequence 
$$K^1(M) \stackrel{i^{\ast}}\longrightarrow K^1(\pM) \stackrel{\delta}\longrightarrow K^0(M,\pM)=\Z \longrightarrow 0$$
by $K^1(X)$, we see that the right column of this diagram is also exact.

It follows from the diagram that $\Indt$ vanishes on both $\Gbox$ and $G^{\p}$.
Both $\alpha_{\p}$ and $\Id\otimes\delta$ are surjective, so $\Indt$ is also surjective.
Finally,
\begin{multline*}
  K^0(\pMX)/G = \im\br{\alpha_{\p}} / \im\br{\alpha_{\p}\circ(\Id\times i)^{\ast}} = \\
	   = \br{K^1(X)\otimes K^1(\pM)} / \br{K^1(X)\otimes K^1(M)} = K^1(X)\otimes\Z,
\end{multline*}
and the quotient map is given by the composition $(\Id\otimes\delta)\circ\alpha_{\p} = \Indt$.
This completes the proof of the proposition.
\endproof

\sub{Special subspaces.}
The following two subspaces of $\Ellt(E)$ will play special role:
\begin{itemize}\upskip
	\item $\Ellp(E)$ consists of all $(A,T)\in\Ellt(E)$ with positive definite $T$.
	\item $\Ellm(E)$ consists of all $(A,T)\in\Ellt(E)$ with negative definite $T$.
\end{itemize}

\begin{prop}\label{prop:Fmp}
Let $\gamma$ be a section of $\Ellt(\E)$. 
Then the following statements hold:
\begin{itemize}\upskip
	\item $\F(\gamma)=0$ if and only if $\gamma$ is a section of $\Ellp(\E)$;
	\item $\F(\gamma)=\EEmY(\gamma)$ if and only if $\gamma$ is a section of $\Ellm(\E)$.
\end{itemize}
\end{prop}

\proof
This follows immediately from the definition of $\F$.
\endproof

Denote by $\Gammapm\Ellt(\E)$ the subspace of $\Gam\Ellt(\E)$
consisting of sections $\gamma$ that can be written in the form 
\begin{equation}\label{eq:Gampm}
  \gamma = \gamma'\oplus\gamma'' \; \text{ with } \;
	\gamma'\in\Gam\Ellp(\E') \text{ and } \gamma''\in\Gam\Ellm(\E'')
\end{equation}
for some orthogonal decomposition $\E\cong\E'\oplus\E''$.

\begin{prop}\label{prop:Fmp2}
The class of $\F(\gamma)$ in $K^0(\pMX)$ lies in $G^{\p}$ 
for every $\gamma\in\Gammapm\Ellt(\E)$.
\end{prop}

\proof
For $\gamma$ defined by \eqref{eq:Gampm}, 
$\F(\gamma) = \F(\gamma'') = \EEmY(\gamma'')$, 
so $[\F(\gamma)]\in G^{\p}$.
\endproof

\sub{Twisting.}
A bundle $\E\in\VectXM$ can be twisted by $W\in\VectX$, 
giving rise to another bundle from $\VectXM$, which we denote by $W\otimes\E$.
If $W$ is a subbundle of a trivial vector bundle $k_X$,
then $W\otimes\E$ is a subbundle of the direct sum of $k$ copies of $\E$,
whose fiber over $x\in X$ is $W_x\otimes\E_x$.

A section $\gamma$ of $\Ellt(\E)$ can be twisted by $W$, 
resulting in the section $1_W\otimes\gamma$ of $\Ellt(W\otimes\E)$.
This construction induces the map $1_W\otimes\colon\Ellt(\E)\to\Ellt(W\otimes\E)$.

For $W\in\VectX$ and $E\in\VectM$ we denote by $W\boxtimes E$
the tensor product $W\otimes\E$, where $\E$ is the trivial bundle over $X$ with the fiber $E$.
For $(A,L)\in\Ellt(E)$ we denote by $1_W\boxtimes(A,L)$ the section $1_W\otimes\gamma$ of $W\boxtimes E$, 
where $\gamma\colon X\to\Ellt(E)$ is the constant map $x\mapsto(A,L)$.

Denote by $\Gammabox \Ellt(\E)$ the subspace of $\Gam\Ellt(\E)$ 
consisting of sections $\gamma$ having the form 
\begin{equation}\label{eq:box}
  \gamma = \bigoplus_{i}{1_{W_i}\boxtimes(A_i,L_i)}	
\end{equation}
for some $(A_i,L_i)\in\Ellt(E_i)$, $E_i\in\VectM$, and $W_i\in\VectX$
with respect to some decomposition of $\E$ into the orthogonal direct sum $\bigoplus_{i}{W_i\boxtimes E_i}$.

\begin{prop}\label{prop:Gbox1}
The class of $\F(\gamma)$ in $K^0(\pMX)$ lies in $G^{\boxtimes}$ 
for every $\gamma\in\Gammabox\Ellt(\E)$.
\end{prop}

\proof
For $\gamma$ defined by formula \eqref{eq:box} we have
$[\F(\gamma)] = \sum_{i}{[W_i\boxtimes F(A_i,L_i)]} \in G^{\boxtimes}$.
\endproof

\sub{Properties of the topological index.}
A continuous map $f\colon X\to Y$ induces the map 
$f^*_{\E}\colon\Gam\Ellt(\E)\to\Gam\Ellt(f^*\E)$ for every $\E\in\Vect_{Y,\,M}$.
On the other hand, $f$ induces the homomorphism $f^*\colon K^1(Y)\to K^1(X)$.
We will use this functoriality to state property (T3) in the following proposition.
 
\begin{prop}\label{prop:prop-indt}
The topological index has the following properties for every $\E,\E'\in\VectXM$: 
\begin{enumerate}\upskip
	\item[(T0)] 
	            The topological index vanishes on $\Gammapm\Ell(\E)$ and $\Gammabox\Ellt(\E)$.
	\item[(T1)] 
	            $\indt(\gamma) = \indt(\gamma')$ if $\gamma$ and $\gamma'$ are homotopic sections of $\Ellt(\E)$.
	\item[(T2)] 
	            $\indt(\gamma\oplus\gamma') = \indt(\gamma) + \indt(\gamma') \in K^0(X)$ 
	            for every section $\gamma$ of $\Ellt(\E)$ and $\gamma'$ of $\Ellt(\E')$.
	\item[(T3)] 
	            $\indt(f^*\gamma) = f^*\indt(\gamma) \in K^1(Y)$ for any section $\gamma$ of $\Ellt(\E)$
							and any continuous map $f\colon Y\to X$.
	\item[(T4)] 
	            $\indt(1_W\otimes\gamma) = [W]\cdot\indt(\gamma)$ for every section $\gamma$ of $\Ellt(\E)$
							and every $W\in\VectX$.
	\item[(T5)] 
	            For a loop $\gamma\colon S^1 \to\Ellt(E)$,	            
							\begin{equation}\label{eq:indt-c1}
								\indt(\gamma) = c_1(\F(\gamma))[\pMS]
							\end{equation}							 
							up to the natural identification $K^1(S^1)\cong\Z$.
	            Here $c_1(\F)$ is the first Chern class of $\F$, 
							$[\pMS]$ is the fundamental class of $\pMS$,
							and $\pM$ is equipped with an orientation in such a way that the pair
							(outward normal to $\pM$, positive tangent vector to $\pM$) has a positive orientation.
\end{enumerate}
\end{prop}

Vanishing of $\indt$ on $\Gammabox\Ellt(\E)$ is a corollary of (T3) and (T4); 
however, we prefer to give this property separately in (T0) for a reason which will be clear later.

\bigskip\proof
(T0).
If $\gamma\in\Gammapm\Ellt(\E)$, then $[\F(\gamma)]\in G^{\p}$ by Proposition \ref{prop:Fmp2}.
If $\gamma\in\Gammabox \Ellt(\E)$, then $[\F(\gamma)]\in G^{\boxtimes}$ by Proposition \ref{prop:Gbox1}.
In both cases Proposition \ref{prop:Indt} implies $\indt(\gamma) = 0$.

(T1). 
If $\gamma$ and $\gamma'$ are homotopic sections of $\Ellt(\E)$,  
then $\F(\gamma)$ and $\F(\gamma')$ are homotopic subbundles of $\EEY$. 
Thus the subbundles $\bra{\F(\gamma)}$ and $\bra{\F(\gamma')}$ of $\bra{\EEY}$ are homotopic,
so they are isomorphic as vector bundles and their classes in $K^0(\pMX)$ coincide.
This implies $\indt(\gamma) = \indt(\gamma')$.

(T2). 
$\F(\gamma\oplus\gamma') = \F(\gamma) \oplus \F(\gamma')$, 
so $[\F(\gamma\oplus\gamma')] = [\F(\gamma)] + [\F(\gamma')]$ in $K^0(\pMX)$.
Applying the homomorphism $\Indt$, we obtain the equality
$\indt(\gamma\oplus\gamma') = \indt(\gamma) + \indt(\gamma')$ in $K^1(X)$.

(T3). 
$\F(f^*\gamma) = f^*\F(\gamma)$, so $[\F(f^*\gamma)] = f^*[\F(\gamma)] \in K^0(Y\times\pM)$.
Since the homomorphism $\Indt\colon K^0(\pMX)\to K^1(X)$ is natural by $X$, we have
$\indt(f^*\gamma) = f^*\indt(\gamma)$.

(T4). 
$\bra{\F(1_W\otimes\gamma)} = W\otimes\bra{\F(\gamma)}$,
so $[\F(1_W\otimes\gamma)] = [W]\cdot[\F(\gamma)]\in K^0(\pMX)$.
Both $\alpha_{\p}\colon K^0(\pMX)\to K^1(X)\otimes K^1(\pM)$ and 
$\Id\otimes\delta\colon K^1(X)\otimes K^1(\pM) \to K^1(X)\otimes\Z$ 
are homomorphisms of $K^0(X)$-modules, 
so their composition $\Indt\colon K^0(\pMX)\to K^1(X)$ is also a homomorphism of $K^0(X)$-modules.
Combining all this together, we get
$\indt(1_W\otimes\gamma) = [W]\cdot\indt(\gamma)$.

(T5). 
It is easy to check that, for $X=S^1$ and up to the natural identification $K^1(S^1)\cong\Z$, 
$\Indt\,[V] = c_1(V)[\pMS]$ for every vector bundle $V$ over $\pMS$.
This implies formula \eqref{eq:indt-c1}
and completes the proof of the proposition.
\endproof

\section{Dirac operators}\label{sec:Dir}

For $k\in\N$ we denote by $k_M$ the trivial vector bundle over $M$ of rank $k$ with the standard Hermitian structure.
Denote by $\kXM\in\VectXM$ the trivial bundle over $X$ with the fiber $k_M$.

\sub{Odd Dirac operators.}
Recall that $A\in\Ell(E)$ is called a Dirac operator if
$\sigma_A(\xi)^2 = \norm{\xi}^2\Id_E$ for all $\xi\in T^{\ast}M$.
We denote by $\Dir(E)$ the subspace of $\Ell(E)$
consisting of all \textit{odd} Dirac operators,
that is, operators having the form 
\begin{equation}\label{eq-Dir-odd}
	A = \matr{ 0 & A^- \\ A^+ & 0}
	\text{ with respect to the chiral decomposition } E = \Ep(A)\oplus\Em(A).
\end{equation}
Denote by $\Dirt(E)$ the subspace of $\Ellt(E)$
consisting of all pairs $(A,L)$ such that $A\in\Dir(E)$.
The following two subspaces of $\Dirt(E)$ will play special role:
$$\Dirp(E) = \Dirt(E)\cap\Ellp(E), \quad \Dirm(E) = \Dirt(E)\cap\Ellm(E).$$
We denote by $\Dirt(\E)$ the subbundle of $\Ellt(\E)$, whose fiber over $x\in X$ is $\Dirt(\E_x)$.
Similarly, denote by $\Dirp(\E)$ and $\Dirm(\E)$ the subbundles of $\Dirt(\E)$, 
whose fibers over $x\in X$ are $\Dirp(\E_x)$ and $\Dirm(\E_x)$, respectively.

\sub{Realization of bundles.}
We will need the following result in our proofs. 

\begin{prop}\label{prop:Fany}
Let $\V\in\VectXM$ and let $\W$ be a subbundle of $\V_{\p}$.
Then there is a section $\gamma$ of $\Dirt(\V\oplus\V)$ such that 
$\EEm(\gamma) = \V\oplus 0$ and $\F(\gamma) = \W$.
In particular, every vector bundle over $\pMX$ is isomorphic to $\bra{\F(\gamma)}$ 
for some $\gamma\colon X\to\Dirt(2k_M))$, $k\in\N$.
\end{prop}

\proof
Let us choose smooth global sections $e_1$, $e_2$ of $TM$ such that
$(e_1(y), e_2(y))$ is a positive oriented frame in $T_y M$ for every $y\in M$.
Choose a smooth unitary connection $\nabla^x$ on each fiber $\V_x$ 
in such a way that $\nabla^x$ continuously depends on $x$ 
with respect to the $C^1$-topology on the space of smooth connections on $\V_x$.
(Such a connection can be constructed using a partition of unity 
subordinated to a finite open covering of $X$ trivializing $\V$.) 
Then $D_x = -i\nabla^x_{e_1} + \nabla^x_{e_2}$ is the Dirac operator acting on sections of $\V_x$ 
and depending continuously on $x$.
Let $D^t_x$ be the operator formally adjoint to $D_x$.
Since the operation of taking formally adjoint operator is a continuous transformation of $\Ell(E)$,
$D^t_x$ is continuous by $x$.
Thus the operator $A_x = \smatr{0 & D_x^t \\ D_x & 0 }$ is an odd self-adjoint Dirac operator 
acting on sections of $\V_x\oplus\V_x$ and depending continuously on $x$.

Let $T_x$ be the self-adjoint automorphism of $\Em(A_x)=\V_{\p,x}\oplus 0$ 
equal to minus the identity on $\W_x$ and to the identity on the orthogonal complement of $\W_x$ in $\V_x$.
Let $L_x$ be the subbundle of $\V_x\oplus\V_x$ corresponding to $T_x$ by formula \eqref{eq:ortL}.
Then $(A_x,L_x)\in\Dirt(\V_x\oplus\V_x)$ and $F(A_x,L_x) = \W_x$.
The section $\gamma\colon x\mapsto (A_x,L_x)$ of $\Dirt(\V\oplus\V)$ 
satisfies conditions $\EEm(\gamma) = \V$ and $\F(\gamma) = \W$,
which proves the first claim of the proposition.

Suppose now that we are given an isomorphism class of a vector bundle over $\pMX$.
We can realize it as a subbundle of a trivial vector bundle $k_{\pMX}$ for some $k\in\N$.
By Proposition \ref{prop:Vany} from the appendix, 
this subbundle is homotopic (and thus isomorphic) to $W = \bra{\W}$ for some subbundle $\W$ of $\kXpM$.
Applying conclusion above to $\V=\kXM$ and $\W$, 
we obtain a section $\gamma$ of $\Dirt(\V\oplus\V)$ such that $\W = \F(\gamma)$. 
Since $\V\oplus\V = 2\kXM$ is trivial, $\gamma$ is just a map from $X$ to $\Dirt(2k_{M})$.
This completes the proof of the proposition.
\endproof

\sub{Image in $K^0(\pMX)$.}
Denote by $\Gammapm\Dirt(\E)$ the subspace of $\Gam\Dirt(\E)$
consisting of sections $\gamma$ that can be written in the form $\gamma = \gamma'\oplus\gamma''$
with $\gamma'\in\Gam\Dirp(\E')$ and $\gamma''\in\Gam\Dirm(\E'')$
for some orthogonal decomposition $\E\cong\E'\oplus\E''$.

\begin{prop}\label{prop:Gpm}
The subgroup $G^{\p}$ of $K^0(\pMX)$ is generated by the classes $[\F(\gamma)]$ 
with $\gamma$ running over $\Gammapm\Dirt(2\kXM)$ and $k$ running over $\N$.
\end{prop}

\proof
The subgroup $G^{\p}$ is generated by the images $j^{\ast}[V]$ with $V\in\Vect_{\MX}$.
By Proposition \ref{prop:Vany}, every such $V$ is isomorphic to $\bra{\V}$ 
for some subbundle $\V$ of $\kXM$ for some (sufficiently large) $k$.
Let $\V'$ be the subbundle of $\kXM$ whose fibers $\V'_x$ are the orthogonal complements 
of fibers $\V_x$ in $k_M$.
By Proposition \ref{prop:Fany}, there are sections $\gamma\in\Gam\Dirt(\V\oplus\V)$ and 
$\gamma'\in\Gam\Dirt(\V'\oplus\V')$ such that 
$\EEm(\gamma) = \V$, $\F(\gamma) = \V_{\p}$, $\EEm(\gamma') = \V'$, and $\F(\gamma') = 0$.
By Proposition \ref{prop:Fmp} $\gamma\in\Gam\Dirm(\V\oplus\V)$ and $\gamma'\in\Gam\Dirp(\V'\oplus\V')$.
Identifying $(\V\oplus\V)\oplus(\V'\oplus\V')$ with $(\V\oplus\V')\oplus(\V\oplus\V') = 2\kXM$,
we identify $\gamma\oplus\gamma'$ with an element of $\Gamma^{\pm}\Dirt(2\kXM)$.
By construction, $\F(\gamma\oplus\gamma') = \V_{\p}\oplus 0$, so 
$j^{\ast}[V] = [\V_{\p}] = [\F(\gamma\oplus\gamma')]$.
This completes the proof of the proposition.
\endproof

\sub{Tensor product.}
Twisting respects Dirac operators and their grading, 
so its restriction to $\Dirt(\E)$ defines the map 
$1_W\otimes\colon\Dirt(\E)\to\Dirt(W\otimes\E)$. 

Denote by $\Gammabox\Dirt(\E)$ the subspace of $\Gammabox\Ellt(\E)$ 
consisting of sections $\gamma$ having the form 
$\gamma = \bigoplus_{i}{1_{W_i}\boxtimes(A_i,L_i)}$	
for some $(A_i,L_i)\in\Dirt(E_i)$, $E_i\in\VectM$, and $W_i\in\VectX$
with respect to some decomposition of $\E$ into the orthogonal direct sum $\bigoplus_{i}{W_i\boxtimes E_i}$.

\begin{prop}\label{prop:Gbox}
The subgroup $G^{\boxtimes}$ of $K^0(\pMX)$ is generated by the classes $[\F(\gamma)]$ 
with $\gamma$ running over $\Gammabox\Dirt(2\kXM)$ and $k$ running over $\N$.
\end{prop}

\proof
The subgroup $G^{\boxtimes}$ is generated by the classes of external tensor products 
$[W\boxtimes V]$ with $W\in\VectX$ and $V\in\VectpM$.
Choose an embedding of $W$ to a trivial vector bundle $n_X$ over $X$,
and let $W'$ be the orthogonal complement of $W$ in $n_X$.
By Proposition \ref{prop:Fany} applied to a one-point base space, 
we can realize $V$ as $F(A,L)$ for some $(A,L)\in\Dirt(2k_M)$, $k\in\N$. 
Chose arbitrary $(A',L')\in\Dirp(2k_M)$.
Then $\gamma = 1_W\boxtimes(A,L)$ is a section of $\Dirt(W\boxtimes 2k_M)$
and $\gamma' = 1_{W'}\boxtimes(A',L')$ is a section of $\Dirp(W'\boxtimes 2k_M)$.
Identifying $W\boxtimes 2k_M \oplus W'\boxtimes 2k_M$ with $(W\oplus W')\boxtimes 2k_M = 2n\kXM$,
we obtain the section $\gamma\oplus\gamma'\in\Gammabox \Dirt(2n\kXM)$ with 
$\bra{\F(\gamma\oplus\gamma')} = (W\boxtimes V) \oplus (W'\boxtimes 0) = W\boxtimes V$.
This completes the proof of the proposition.
\endproof

\sub{Surjectivity of the topological index.}

\begin{prop}\label{prop:sur-indt}
For every $\mu\in K^1(X)$ there are $k\in\N$ and $\gamma\colon X\to\Dirt(2k_M)$ such that $\mu=\indt(\gamma)$.
\end{prop}

\proof
By Proposition \ref{prop:Indt} the homomorphism $\Indt\colon K^0(\pMX)\to K^1(X)$ is surjective,
so $\mu = \Indt\lambda$ for some $\lambda\in K^0(\pMX)$.
We can realize $\lambda$ as $[V]-[n_{\pMX}]$ for some vector bundle $V$ over $\pMX$ and $n\in\N$. 
By Proposition \ref{prop:Fany} $V$ is isomorphic to $\bra{\F(\gamma)}$ for some $\gamma\colon X\to\Dirt(2k_M)$.
The trivial vector bundle $n_{\pMX}$ is the restriction of $n_{\MX}$ to $\pMX$, so
$[n_{\pMX}]\in G^{\p}\subset\Ker\Indt$. 
Combining all this, we obtain
$$\indt(\gamma) = \Indt\,[V] = \Indt\,[V] - \Indt\,[n_{\pMX}] = \Indt\lambda = \mu.$$
This completes the proof of the proposition.
\endproof

\section{Universality of the topological index}\label{sec:uni-top}

\upskip

\sub{Homotopies that fix operators.}
In this section we will deal with those deformations of sections of $\Ellt(\E)$ 
that fix an operator family $(A_x)$ and change only boundary conditions $(L_x)$. 

Let us fix an odd Dirac operator $D\in\Dir(2_M)$.
Denote by $\delta^+$, resp. $\delta^-$ the constant map from $X$ to $(D,\Id)\in\Dirp(2_M)$, resp. $(D,-\Id)\in\Dirm(2_M)$.
We denote by $k\delta^+$, resp. $k\delta^-$ the direct sum of $k$ copies of $\delta^+$, resp. $\delta^-$.

\begin{prop}\label{prop:Fhom}
Let $\gamma\colon x\mapsto(A_x,L_x)$ and $\gamma'\colon x\mapsto(A_x,L'_x)$ 
be sections of $\Ellt(\E)$ differing only by boundary conditions.
Then the following statements hold.
\begin{enumerate}\upskip
	\item If $\bra{\F(\gamma)}$ and $\bra{\F(\gamma')}$ are homotopic subbundles of $\bra{\EEmY(\gamma)}$,
then $\gamma$ and $\gamma'$ are homotopic sections of $\Ellt(\E)$.
	\item If $\bra{\F(\gamma)}$ and $\bra{\F(\gamma')}$ are isomorphic as vector bundles,
then the sections $\gamma\oplus k\delta^+$ and $\gamma'\oplus k\delta^+$
of $\Ellt(\E\oplus 2\kXM)$ are homotopic for some $k\in\N$.
	\item If $[\F(\gamma)] = [\F(\gamma')] \in K^0(\pMX)$,
then the sections $\gamma\oplus l\delta^-\oplus k\delta^+$ and $\gamma'\oplus l\delta^-\oplus k\delta^+$
of $\Ellt(\E\oplus 2l_{X,\,M}\oplus 2\kXM)$ are homotopic for some $l,k\in\N$.
\end{enumerate}
\end{prop}

\proof
Recall that $\EEmY(\gamma)$ depends only on operators, 
so $\EEmY(\gamma) = \EEmY(\gamma')$; denote it by $\EEmY$.
Denote $\F = \F(\gamma)$ and $\F' = \F(\gamma')$. 

1. Let $\A\colon x\mapsto A_x$ be the corresponding section of $\Ell(\E)$.
Denote by $\L(\A)\subset\Gam\Ellt(\E)$ the space of all lifts of $\A$ to sections of $\Ellt(\E)$.
Denote by $\L^u(\A)$ the subspace of $\L(\A)$ consisting of sections $(A_x,T_x)$ 
such that the self-adjoint automorphisms $T_x$ is unitary for every $x\in X$.
The subspace $\L^u(\A)$ is a strong deformation retract of $\L(\A)$, with the retraction given by the formula
$h_s(A_x,T_x) = (A_x, (1-s+s|T_x|\inv)T_x)$.
Since $h_s$ preserves $\F$,
it is sufficient to prove the first claim of the proposition for $\gamma,\gamma'\in\L^u(\A)$.

Suppose that $\bra{\F}$ and $\bra{\F'}$ are homotopic subbundles of $\bra{\EEmY}$.
Then $\F$ and $\F'$ are homotopic subbundles of $\EEmY$ by Proposition \ref{prop:Vhom} from the appendix.
An element $\gamma\in\L^u(\A)$ is uniquely defined by the subbundle $\F(\gamma)$ of $\EEmY(\gamma)$.
Hence a homotopy between $\F$ and $\F'$ defines the path in $\L^u(\A)\subset\Gam\Ellt(\E)$
connecting $\gamma$ with $\gamma'$.

2. If $\bra{\F}$ and $\bra{\F'}$ are isomorphic as vector bundles,
then they are homotopic as subbundles of $\bra{\EEmY}\oplus k_{\pMX}$ for $k$ large enough.
Thus the sections $\gamma\oplus k\delta^+$ and $\gamma'\oplus k\delta^+$ of $\Ellt(\E\oplus 2\kXM)$
satisfy conditions of the first claim of the proposition and therefore are homotopic.

3. The equality $[\F] = [\F']$ implies that
the vector bundles $\bra{\F}$ and $\bra{\F'}$ are stably isomorphic,
that is, $\bra{\F}\oplus l_{\pMX} = \bra{\F(\gamma\oplus l\delta^-)}$ 
and $\bra{\F'}\oplus l_{\pMX} = \bra{\F(\gamma'\oplus l\delta^-)}$ are isomorphic for some integer $l$.
It remains to apply the second part of the proposition to the sections
$\gamma\oplus l\delta^-$ and $\gamma'\oplus l\delta^-$ of $\Ellt(\E\oplus 2l_{X,\,M})$.
\endproof

\sub{The case of different operators.}
For a section $\gamma\colon x\mapsto(A_x,T_x)$ of $\Ellt(\E)$ 
we denote by $\gamma^+$ the section of $\Ellp(\E)$ given by the rule $x\mapsto (A_x,\Id)$. 

\begin{prop}\label{prop:F-iso2}
Let $\gamma_i$ be a section of $\Ellt(\E_i)$, $\E_1,\E_2\in\VectXM$, $i=1,2$.
Suppose that $[\F(\gamma_1)] = [\F(\gamma_2)] \in K^0(\pMX)$.
Then the sections 
$\gamma_1\oplus\gamma_2^+\oplus l\delta^-\oplus k\delta^+$ and 
$\gamma_1^+\oplus\gamma_2\oplus l\delta^-\oplus k\delta^+$
of $\Ellt(\E_1\oplus\E_2\oplus 2l_{X,\,M}\oplus 2\kXM)$ are homotopic 
for $l$, $k$ large enough.
\end{prop}

\proof
The sections $\gamma'_1 = \gamma_1\oplus\gamma_2^+$ and $\gamma'_2 = \gamma_1^+\oplus\gamma_2$
of $\Ellt(\E_1\oplus\E_2)$ differ only by boundary conditions 
and thus fall within the framework of Proposition \ref{prop:Fhom}.
By Proposition \ref{prop:Fmp} $\F(\gamma'_i) = \F(\gamma_i)$.
It remains to apply the third part of Proposition \ref{prop:Fhom} to $\gamma'_1$ and $\gamma'_2$. 
\endproof

\sub{Commutativity.}
The direct sum of operators is a non-commutative operation. 
However, it is commutative up to homotopy, as the following proposition shows.

\begin{prop}\label{prop:comm1}
Let $f\colon X\to\Ellt(2k_M)$, $g\colon X\to\Ellt(2l_M)$ be continuous maps.
Then $f\oplus g$ and $g\oplus f$ are homotopic as maps from $X$ to $\Ellt((2k+2l)_M)$.
\end{prop}

\proof
Let $J_1$ be the unitary automorphism of $\CC^{2k+2l}$
given by the formula $u\oplus v \mapsto v\oplus u$ for $u\in\CC^{2k}$, $v\in\CC^{2l}$.
Let us choose a path $J\colon[0,1]\to \U(\CC^{2k+2l})$ 
connecting $J_0=\Id$ with $J_1$.
	Denote by $\tilde{J}_s$ the unitary bundle automorphism of $(2k+2l)_M$ induced by $J_s$.
	Then the map $h\colon[0,1]\times X \to\Ellt((2k+2l)_M)$ defined by the formula 
	$h_s(x) = \tilde{J}_s(f(x)\oplus g(x))$ 
	gives a desired homotopy between $f\oplus g$ and $g\oplus f$.
\endproof

\sub{Universality of the topological index.}
Now we are ready to state our first universality result.

\begin{thm}\label{thm:uni-top}
Let $\gamma_i$ be a section of $\Ellt(\E_i)$, $\E_1,\E_2\in\VectXM$, $i=1,2$.
		Then the following two conditions are equivalent:
\begin{enumerate}\upskip
	\item $\indt(\gamma_1) = \indt(\gamma_2)$.
	\item There are $k,l\in\N$ and sections $\beta_i^{\pm}\in\Gammapm\Dirt(2\kXM)$,
	      $\beta_i^{\boxtimes}\in\Gammabox\Dirt(2l_{X,M})$ 
	      	such that 	
	\begin{equation}\label{eq:sum}
		\gamma_1\oplus\gamma_2^+\oplus\beta_1^{\pm}\oplus\beta_1^{\boxtimes}  \quad \text{and} \quad
		\gamma_1^+\oplus\gamma_2\oplus\beta_2^{\pm}\oplus\beta_2^{\boxtimes}
	\end{equation}
	are homotopic sections of $\Ellt(\E_1\oplus\E_2\oplus 2\kXM\oplus 2l_{X,M})$.
\end{enumerate}
\end{thm}

\proof
($2\Rightarrow 1$) follows immediately from properties (T0--T2) of the topological index.

Let us prove ($1\Rightarrow 2$).
Suppose that $\indt(\gamma_1) = \indt(\gamma_2)$.
Then $\Indt(\lambda_1-\lambda_2) = 0$ for $\lambda_i = [\F(\gamma_i)]\in K^0(\pMX)$.
Proposition \ref{prop:Indt} implies that $\lambda_1-\lambda_2 = \lambda^{\p} + \lambda^{\boxtimes}$ 
for some $\lambda^{\p}\in G^{\p}$ and $\lambda^{\boxtimes}\in G^{\boxtimes}$.

By Proposition \ref{prop:Gpm} $\lambda^{\p} = [\F(\beta^{\p}_2)] - [\F(\beta^{\p}_1)]$
for some $\beta^{\p}_1,\beta^{\p}_2\in\Gammapm\Dirt(2n_{X,M})$ 
(one can equate the ranks of corresponding trivial bundles
by adding several copies of $\delta^+$ if needed).
Similarly, by Proposition \ref{prop:Gbox} 
$\lambda^{\boxtimes} = [\F(\beta^{\boxtimes}_2)] - [\F(\beta^{\boxtimes}_1)]$
for some $\beta^{\boxtimes}_1,\beta^{\boxtimes}_2\in\Gammabox\Dirt(2l_{X,M})$
(one can equate the ranks of corresponding trivial bundles 
by increasing the ranks of ambient trivial bundles for $V$ and $W$ in construction of $\beta^{\boxtimes}_i$ if needed; 
see the proof of Proposition \ref{prop:Gbox}).
Combining all this, we obtain 
$$\brr{\F\br{\gamma_1\oplus\beta_1^{\boxtimes}\oplus\beta_1^{\p}}} = 
  \brr{\F\br{\gamma_2\oplus\beta_2^{\boxtimes}\oplus\beta_2^{\p}}}.$$
Adding sections of $\Ellp(\E_i\oplus 2l_{X,\,M}\oplus 2n_{X,\,M})$ to the sections on both sides of this equality, 
we obtain
$$\brr{\F\br{\gamma_1\oplus\gamma_2^+ \oplus(\beta_1^{\boxtimes}\oplus\beta_1^{\p}) \oplus(\beta_2^{\boxtimes}\oplus\beta_2^{\p})^+}} = 
  \brr{\F\br{\gamma_1^+\oplus\gamma_2 \oplus(\beta_1^{\boxtimes}\oplus\beta_1^{\p})^+ \oplus(\beta_2^{\boxtimes}\oplus\beta_2^{\p})}}.$$
The third part of Proposition \ref{prop:Fhom}	implies that 	
$$\gamma_1\oplus\gamma_2^+ \oplus(\beta_1^{\boxtimes}\oplus\beta_1^{\p}) 
  \oplus(\beta_2^{\boxtimes}\oplus\beta_2^{\p})^+ \oplus s\delta^-\oplus t\delta^+$$
and
$$\gamma_1^+\oplus\gamma_2 \oplus(\beta_1^{\boxtimes}\oplus\beta_1^{\p})^+ 
  \oplus(\beta_2^{\boxtimes}\oplus\beta_2^{\p}) \oplus s\delta^-\oplus t\delta^+$$	
are homotopic for some integers $s$, $t$.
Using Proposition \ref{prop:comm1} to rearrange terms,
taking $k = 2n+l+s+t$, and defining $\beta_i^{\pm}\in\Gammapm\Dirt(2\kXM) $ by the formula
\[\beta_i^{\pm} = \beta_i^{\p}\oplus(\beta_j^{\p}\oplus\beta_j^{\boxtimes})^+\oplus s\delta^-\oplus t\delta^+ 
   \; \text{ for } \set{i,j} = \set{1,2},\]
we obtain the second condition of the theorem. 
\endproof

\sub{Universality for families.}
Our next goal is to describe invariants of families of elliptic operators satisfying 
the same properties as the topological index.
Let $\Phi(\gamma)$ be such an invariant.
We start with the first three properties (T0-T2) of the topological index: 
\begin{enumerate}\upskip
	\item[(\Epm)] $\Phi$ vanishes on $\Gammapm\Ellt(\E)$.
	\item[(\Ebox)] $\Phi$ vanishes on $\Gammabox\Ellt(\E)$.
	\item[(E1)] $\Phi(\gamma) = \Phi(\gamma')$ if $\gamma$ and $\gamma'$ are homotopic sections of $\Ellt(\E)$.
	\item[(E2)] $\Phi(\gamma\oplus\gamma') = \Phi(\gamma) + \Phi(\gamma')$ 
	    for every section $\gamma$ of $\Ellt(\E)$ and $\gamma'$ of $\Ellt(\E')$.
\end{enumerate}

Let $\Ve$ be a subclass of $\VectXM$ satisfying the following condition:
\begin{multline}\label{eq:Ve}
	\Ve \text{ is closed under direct sums} \\
	\text{and contains the trivial bundle } 2\kXM \text{ for every } k\in\N.
\end{multline}
In particular, $\Ve$ can coincide with the whole $\VectXM$.

\begin{thm}\label{thm:top-fam}
Let $X$ be a compact space and $\Lambda$ be a commutative monoid.
Suppose that we associate an element $\Phi(\gamma)\in\Lambda$
with every section $\gamma$ of $\Ellt(\E)$ for every $\E\in\Ve$.
Then the following two conditions are equivalent:
\begin{enumerate}\upskip
	\item $\Phi$ satisfies properties (\Epm, \Ebox, E1, E2) for all $\E,\E'\in\Ve$;
	\item $\Phi$ has the form $\Phi(\gamma) = \vartheta(\indt(\gamma))$
        for some (unique) monoid homomorphism $\vartheta\colon K^1(X)\to\Lambda$.
\end{enumerate}
\end{thm}

\begin{rem}\label{rem:top-fam}
In the case $\Ve=\VectXM$,
the property (\Epm) in the statement of this theorem can be replaced 
by the property (\Tpm) from the Introduction, 
namely vanishing of $\Phi$ on sections of $\Ellp(\E)$ and $\Ellm(\E)$. 
Indeed, a section from $\Gammapm\Ellt(\E)$ is a sum of sections of $\Ellp(\E')$ and $\Ellm(\E'')$ for some $\E'$ and $\E''$, 
so (\Tpm) together with (E2) implies (\Epm).
Similarly, (\Ebox) can be replaced by the property (\Tbox) from the Introduction, 
namely vanishing of $\Phi$ on sections having the form $1_W\boxtimes(A,L)$.
Therefore, for $\Ve=\VectXM$ Theorem \ref{thm:top-fam} takes the form of Theorem \ref{thm:top-fam0}.
\end{rem}

\proof
($2\Rightarrow 1$) follows immediately from properties (T0--T2) of the topological index.

Let us prove ($1\Rightarrow 2$). 
We show first that 
\begin{equation}\label{eq:indtPhi}
	\indt(\gamma_1) = \indt(\gamma_2) \quad \text{implies} \quad \Phi(\gamma_1) = \Phi(\gamma_2)
\end{equation}
for all $\gamma_i\in\Gam\Ellt(\E_i)$, $\E_1,\E_2\in\Ve$.
Indeed, if $\indt(\gamma_1) = \indt(\gamma_2)$, 
then by Theorem \ref{thm:uni-top} the sections \eqref{eq:sum}
are homotopic for some $k,l\in\N$, $\beta_i^{\pm}\in\Gammapm\Dirt(2\kXM)$, 
and $\beta_i^{\boxtimes}\in\Gammabox\Dirt(2l_{X,M})$.
Properties (E1) and (E2) then imply
\[\Phi(\gamma_1)+\Phi(\gamma_2^+)+\Phi(\beta_1^{\pm})+\Phi(\beta_1^{\boxtimes}) = 
  \Phi(\gamma_2)+\Phi(\gamma_1^+)+\Phi(\beta_2^{\pm})+\Phi(\beta_2^{\boxtimes}).\]
(\Epm) implies $\Phi(\gamma_i^+) = \Phi(\beta_i^{\pm}) = 0$, while
(\Ebox) implies $\Phi(\beta_i^{\boxtimes}) = 0$. 
Thus we obtain $\Phi(\gamma_1) = \Phi(\gamma_2)$, which proves \eqref{eq:indtPhi}.

Next we define the homomorphism $\vartheta\colon K^1(X)\to\Lambda$.
Let $\mu$ be an arbitrary element of $K^1(X)$.
By Proposition \ref{prop:sur-indt}
there exist $k\in\N$ and a section $\beta$ of $\Dirt(2\kXM)$ such that $\mu = \indt(\beta)$.
In order to satisfy condition (2) of the theorem we have to put $\vartheta(\mu) = \Phi(\beta)$.
The correctness of this definition follows from \eqref{eq:indtPhi}.

Let now $\gamma$ be an arbitrary section of $\Ellt(\E)$ and $\mu = \indt(\gamma)$. 
By definition above $\vartheta(\mu) = \Phi(\beta)$ for some $\beta$ such that $\mu = \indt(\beta)$.
Then $\indt(\gamma) = \mu = \indt(\beta)$, 
so \eqref{eq:indtPhi} implies $\Phi(\gamma) = \Phi(\beta) = \vartheta(\mu) = \vartheta(\indt(\gamma))$.
This completes the proof of the theorem.
\endproof

\sub{Universality for families: functoriality and twisting.}
Our next goal is to describe
families $\Phi=(\Phi_X)$ of $K^1(X)$-valued invariants 
satisfying two more properties in addition to (\Epm, \Ebox, E1, E2):
\begin{enumerate}\upskip
	\item[(E3)] 
	            $\Phi_Y(f^*_{\E}\gamma) = f^*\Phi_X(\gamma) \in K^1(Y)$ 
							for every section $\gamma$ of $\Ellt(\E)$ and every continuous map $f\colon Y\to X$.
	\item[(E4)] 
	            $\Phi_X(1_W\otimes\gamma) = [W]\cdot\Phi_X(\gamma)$ for every section $\gamma$ of $\Ellt(\E)$
							and every $W\in\VectX$.
\end{enumerate}

\begin{thm}\label{thm:top-fam2}
Suppose that we associate an element $\Phi_X(\gamma)\in K^1(X)$
with every section $\gamma$ of $\Ellt(\E)$ 
for every compact topological space $X$ and every $\E\in\VectXM$.
Then the following two conditions are equivalent:
\begin{enumerate}\upskip
	\item The family $\Phi = (\Phi_X)$ satisfies properties (\Epm, \Ebox,  E1--E4) for all $\E,\E'\in\VectXM$;
	\item There is an integer $m$ such that $\Phi$ has the form $\Phi_X = m\cdot\indt$.	      
\end{enumerate}\upskip
\end{thm}

\upskip
\begin{rem}
As well as in Remark \ref{rem:top-fam},
the property (\Epm) in the statement of this theorem can be replaced by (\Tpm)
and (\Ebox) can be replaced by (\Tbox).
\end{rem}

\proof
($2\Rightarrow 1$) follows immediately from properties (T0-T4) of the topological index.

Let us prove ($1\Rightarrow 2$). 
By Theorem \ref{thm:top-fam}, 
for every compact space $X$ there is a homomorphism 
\begin{equation}\label{eq:indtX}
  \vartheta_X\colon K^1(X)\to K^1(X) \; \text{ such that } \; \Phi_X(\gamma) = \vartheta_X(\indt(\gamma))
\end{equation}
for every $\E\in\VectXM$ and every section $\gamma$ of $\Ellt(\E)$.
Moreover, such a homomorphism $\vartheta_X$ is unique.

Let $f\colon Y\to X$ be a continuous map and $\mu\in K^1(X)$. 
By property (T5) of the topological index $\mu=\indt(\gamma)$ 
for some $\gamma\in\Gam\Ellt(\E)$, $\E\in\VectXM$.
By (T3) $\indt(f^*_{\E}\gamma) = f^*\indt(\gamma)$ and by (E3) $\Phi_Y(f^*_{\E}\gamma) = f^*\Phi_X(\gamma)$.
Substituting this to \eqref{eq:indtX}, we obtain 
\begin{multline*}
  \vartheta_Y(f^*\mu) = \vartheta_Y(f^*\indt(\gamma)) = \vartheta_Y(\indt(f^*_{\E}\gamma)) 
	= \Phi_Y(f^*_{\E}\gamma) = f^*\Phi_X(\gamma) \\
	= f^*\vartheta_X\indt(\gamma) = f^*\vartheta_X(\mu).
\end{multline*}
Thus the family $(\vartheta_X)$ defines a natural transformation $\vartheta$ of the functor $X\mapsto K^1(X)$ to itself.

Similarly, (T4) and (E4) imply that $\vartheta$ respects the $K^0(\cdot)$-module structure on $K^1(\cdot)$,
that is, $\vartheta_X(\lambda\mu) = \lambda\vartheta_X(\mu)$ 
for every compact space $X$ and every $\lambda\in K^0(X)$, $\mu\in K^1(X)$.

We show in Proposition \ref{prop:K1} of the Appendix that 
the only natural transformations satisfying this property are multiplications by an integer.
Hence, there is an integer $m$ such that $\vartheta_X(\mu)=m\mu$ 
for every $X$ and every $\mu\in K^1(X)$.
Substituting this identity to \eqref{eq:indtX}, we obtain the second condition of the theorem.
\endproof

\sub{The semigroup of elliptic operators.}
The disjoint union \[ \Ellt_M := \coprod_{k\in\N}\Ellt(2k_M) \]
has the natural structure of a (non-commutative) graded topological semigroup,
with the grading by $k$ and the semigroup operation given by the direct sum of operators and boundary conditions.
We denote by $\EllXM$ the trivial bundle over $X$ with the fiber $\Ellt_M$ and by
$$\Gam\EllXM = C(X,\Ellt_M)$$ 
the topological semigroup of its sections, with the compact-open topology.

We will use the following two special subsemigroups of $\Gam\EllXM$:
\[
\Gammapm\EllXM = \coprod_{k\in\N}\Gammapm\Ellt(2\kXM) \quad \text{and} \quad
\Gammabox\EllXM = \coprod_{k\in\N}\Gammabox\Ellt(2\kXM).
\]
The subsemigroup of $\Gam\EllXM$ spanned by $\Gammapm\EllXM$ and $\Gammabox\EllXM$
will play a special role; we denote it by $\Gammapmbox\EllXM$.

\sub{The homotopy classes.}
The set $\pi_0(\Gam\EllXM) = [X,\Ellt_M]$ of homotopy classes of maps from $X$ to $\Ellt_M$ 
has the induced semigroup structure. 

\begin{prop}\label{prop:comm}
  The semigroup $[X,\Ellt_M]$ is commutative for any topological space $X$.
\end{prop}

\proof
  Let $f,g\colon X\to\Ellt_M$ be continuous maps.
  For every $k,l\in\N$ the inverse images $f\inv(\Ellt(2k_M))$ and $g\inv(\Ellt(2l_M))$
	are open and closed in $X$, so their intersection $X_{k,l}$ is also open and closed.
	By Proposition \ref{prop:comm1} the restrictions of $f\oplus g$ and $g\oplus f$ to $X_{k,l}$ 
	are homotopic as maps from $X_{k,l}$ to $\Ellt((2k+2l)_M)$
	(the proof of Proposition \ref{prop:comm1} does not use compactness of $X$ 
	and works as well for arbitrary topological space).
	Since $X$ is the disjoint union of $X_{k,l}$, this implies that 
	$f\oplus g$ and $g\oplus f$ are homotopic as maps from $X$ to $\Ellt_M$.
	Therefore, the classes of $f\oplus g$ and $g\oplus f$ in $[X,\Ellt_M]$ coincide,
	so $[X,\Ellt_M]$ is commutative.
	This completes the proof of the proposition.
\endproof

\sub{The topological index as a homomorphism.}
A continuous map $\gamma\colon X\to\Ellt_M$ defines the partition of $X$ by subsets $X_k$,
where $X_k$ consists of points $X$ such that $\gamma(x)$ has the grading $k$.
Since the grading is continuous, all $X_k$ are open-and-closed subsets of $X$.
Since $X$ is compact, all but a finite number of $X_k$ are empty, so this partition is finite.
The restriction of $\gamma$ to $X_k$ takes values in $\Ellt(2k_M)$,
so $\gamma$ can be identified with a section of $\Ellt(\E_{\gamma})$, 
where $\E_{\gamma}\in\VectXM$ is the bundle whose restriction to $X_k$ is 
the trivial bundle over $X_k$ with the fiber $2k_M$.
Thus the topological index of $\gamma$ is well defined.

Since the topological index is additive with respect to direct sums, 
it defines the monoid homomorphism $\ind_t\colon C(X,\Ellt_M)\to K^1(X)$.
Since the topological index is homotopy invariant, 
this homomorphism factors through the projection $C(X,\Ellt_M)\to[X;\Ellt_M]$.

The inclusion $\Gammapmbox\EllXM \hookto \Gam\EllXM$ induces the homomorphism 
$$\pi_0(\Gammapmbox\EllXM) \to \pi_0(\Gam\EllXM) = [X;\Ellt_M];$$ 
we denote its image by $[X;\Ellt_M]\pmbox$.

Since the topological index vanishes on $\Gammapm\EllXM$ and $\Gammabox\EllXM$,
it factors through the quotient $[X;\Ellt_M] / [X;\Ellt_M]\pmbox$.
In other words, there exists a monoid homomorphism 
$$\kappa_t\colon [X;\Ellt_M] / [X;\Ellt_M]\pmbox \to K^1(X)$$ 
such that the following diagram is commutative:
\begin{equation}\label{diag:kappaD}							
	\begin{tikzcd}
		C(X;\Ellt_M) \arrow{r} \arrow[swap]{drr}{\indt} &
		{[X;\Ellt_M]}  \arrow{r} & 
		{[X;\Ellt_M] / [X;\Ellt_M]\pmbox} \arrow{d}{\kappa_t} 
		\\
		 &  &  K^1(X) 
	\end{tikzcd}
\end{equation}

\begin{thm}\label{thm:top-maps2}
Let $X$ be a compact topological space.
Then $[X,\Ellt_M] / [X;\Ellt_M]\pmbox$ is an Abelian group isomorphic to $K^1(X)$,
with an isomorphism given by $\kappa_t$.
\end{thm}

\upskip
Note that, for any given $k$, the restriction of $\kappa_t$ to a given rank,
$$[X,\Ellt(2\kXM)] / [X;\Ellt(2\kXM)]\pmbox \to K^1(X),$$
in general is neither injective nor surjective, 
so we need to take the direct sum for all the ranks to obtain universality. 

\medskip

\proof
Denote the commutative monoid $[X,\Ellt_M]/[X;\Ellt_M]\pmbox$ by $\Lambda$
and the composition of horizontal arrows on diagram \eqref{diag:kappaD} by $\Phi$, so that $\indt = \kappa_t\circ\Phi$.
By definition, $\Phi$ is additive, homotopy invariant, surjective, 
and vanishes on both $\Gammapm\EllXM$ and $\Gammabox\EllXM$.

Suppose first that $X$ is connected.
Then $[X,\Ellt_M] = \coprod_k[X,\Ellt(2k_M)]$, 
so $\Phi$ and $\Lambda$ satisfy the first condition of Theorem \ref{thm:top-fam} with $\Ve=\set{2\kXM}$.
Thus $\Phi = \vartheta\circ\indt$ for some monoid homomorphism $\vartheta\colon K^1(X)\to\Lambda$.
By Proposition \ref{prop:prop-indt} the topological index is surjective.
Thus $\kappa_t$ and $\vartheta$ are mutually inverse and $\kappa_t$ is an isomorphism. 
This completes the proof of the theorem in the case of connected $X$.

In the general case we need to extend the set $\set{2\kXM}_{k\in\N}$ of trivial bundles.
Let $\Ve$ be the set of all bundles $\E_{\gamma}$ with $\gamma\in\Gam\EllXM$.
An element $\E$ of $\Ve$ is defined by a partition of $X$ by open-and-closed subsets $X_k$, $k\in\N$, 
such that all but a finite number of $X_k$ are empty.
For such a partition, $\E$ is defined as the disjoint union of trivial bundles $2k_{X_k,\,M}$.
A continuous map from $X$ to $\EllXM$ is nothing else than a section of a bundle $\Ellt(\E)$ with $\E\in\Ve$.
Obviously, $\Ve$ is closed under direct sums and contains all trivial bundles $2\kXM$.
Hence the triple $(\Ve, \Phi, \Lambda)$ satisfies the first condition of Theorem \ref{thm:top-fam},
and therefore $\Phi = \vartheta\circ\indt$ for some monoid homomorphism $\vartheta\colon K^1(X)\to\Lambda$.
Taking into account that both $\Phi$ and $\indt$ are surjective, 
we see that $\kappa_t$ and $\vartheta$ are mutually inverse and thus $\kappa_t$ is an isomorphism. 
This completes the proof of the theorem.
\endproof

\section{Deformation retraction}\label{sec:retr}

\begin{prop}\label{prop:retr2}
The natural embedding $\Dir(\E)\hookto\Ell(\E)$ is a bundle homotopy equivalence for every $\E\in\VectXM$.
Moreover, there exists a fiberwise deformation retraction $h$ 
of $\Ell(\E)$ onto a subbundle of $\Dir(\E)$
having the following properties
for every $s\in[0,1]$, $A\in\Ell(\E_x)$, and $A_s=h_s(A)$:
\begin{enumerate}\upskip
	\item[(1)] $\Em(A_s)=\Em(A)$.
	\item[(2)] The symbol of $A_s$ depends only on $s$ and the symbol $\sigma_A$ of $A$. 	           
	\item[(3)] The map $h'_s\colon\sigma_{A}\mapsto\sigma_{A_s}$ defined by (2)
						is $U(\E_x)$-equivariant.
	\item[(4)] If $A\in\Dir(\E_x)$, then $\sigma_{A_s} = \sigma_{A}$.
\end{enumerate}\upskip
\end{prop}

\upskip
In the case of one-point space $X$ this result was proved in \cite[Proposition 9.5]{Pr17}.
We will use it to construct such a deformation retraction for an arbitrary compact space $X$.

\medskip
\proof
Let $(X^i)$ be a finite open covering of $X$ such that the restrictions of $\E$ to $X^i$ are trivial.
Choose trivializations $f^i\colon \restr{\E}{X^i}\to E^i\times X^i$.
For $x\in X^i$, denote by $f^{i}_x\in U(\E_x,E^i)$ the isomorphism of the fibers given by $f^i$.
The homeomorphism $\Ell(\E_x) \to \Ell(E^i)$ induced by $f^{i}_x$
we will also denote by $f^{i}_x$.

Choose a partition of unity $(\rho_i)$, $\rho_i \in C(X^i, C^{\infty,1}(M))$, subordinated to the covering $(X^i)$.
Let $h^i\colon[0,1]\times\Ell(E^i)\to\Ell(E^i)$ be a deformation retraction 
of $\Ell(E^i)$ onto a subspace of $\Dir(E^i)$ satisfying the conditions of \cite[Proposition 9.5]{Pr17}. 

For $x\in X^i$ and $A\in\Ell(\E_x)$, 
we define an element $A^{i}_{s}$ of $\Ell(\E_x)$ by the formula 
$f^{i}_x \br{ A^{i}_{s}} = h^i_s\br{f^{i}_x(A)}$.
From \cite[Proposition 9.5]{Pr17} we obtain the following: 
\begin{enumerate}\upskip
	\item[(a)] $A^{i}_{0} = A$ and $A^{i}_{1}\in\Dir(\E_x)$ for every $i$.
	\item[(b)] The symbol of $A^{i}_{s}$ depends only on $s$ and the symbol $\sigma$ of $A$
             and is independent of $i$; denote it by $\sigma_{s}$.
	\item[(c)] The map $\sigma\mapsto\sigma_s$ defined by (b) is $U(\E_x)$-equivariant.
	\item[(d)] $\Em(\sigma_s)=\Em(\sigma)$.
	\item[(e)] If $A\in\Dir(\E_x)$, then $\sigma_{s} = \sigma$ for all $s\in[0,1]$. 
	\item[(f)] If If $A,B\in\Ell(\E_x)$ and the symbols of $A^i_1$ and $B^i_1$ coincide, then $A^i_1=B^i_1$.
\end{enumerate}
We claim that the bundle map $h\colon[0,1]\times\Ell(\E)\to\Ell(\E)$ 
defined by the formula
\begin{equation}\label{eq:Hs}
  h_s(A) = \sum_i\rho_i(x)A^{i}_{s}	\; \text{ for } A\in\Ell(\E_x)
\end{equation}
is a desired deformation retraction.
The rest of the proof is devoted to the verification of this claim.

First note that (a) implies $h_0=\Id$.
A convex combination of self-adjoint elliptic operators with the symbol $\sigma_s$ is again 
a self-adjoint elliptic operator with the symbol $\sigma_s$, 
so (b) implies $\sigma_{A_s} = \sigma_s$ and $A_{s}\in\Ell(\E_x)$.
(c) implies condition (3) of the proposition, 
(e) implies (4), and (d) implies (1). 

The chiral decomposition of an odd Dirac operator $A^{i}_{1}$
is defined by its symbol $\sigma_{1}$ and hence is independent of $i$,
so (a) and (b) imply $\im h_1\subset\Dir(\E)$.

Suppose that $A\in\im h_1$, that is, $A=B_1$ for some $B\in\Ell(\E_x)$.
Then $A\in\Dir(\E_x)$, and (e) implies $\sigma_{A_1} = \sigma_A = \sigma_{B_1}$.
Hence the symbols of $A_1^i$ and $B_1^i$ coincide,
and (f) implies $A_1^i = B_1^i$.
Substituting this to \eqref{eq:Hs}, we obtain $A_1=B_1$, that is, $h_1(A) = A$.
Thus the restriction of $h_1$ on its image is the identity.

It remains to prove the homotopy equivalence part.
Let $A\in\Dir(\E_x)$. 
Then $A_1 = h_1(A)$ also lies in $\Dir(\E_x)$, 
but $A_s$ is not necessarily odd for $s\in(0,1)$, so we should change a homotopy a little.
Since the symbols of $A_1$ and $A$ coincide, 
the formula $h'_s(A) = (1-s)A+sA_1$ defines a continuous bundle map
$h'\colon[0,1]\times\Dir(\E)\to\Dir(\E)$ such that $h'_0=\Id$ and $h'_1=h_1$.
It follows that the restriction of $h_1$ to $\Dir(\E)$ and the identity map $\Id_{\Dir(\E)}$
are homotopic as bundle maps from $\Dir(\E)$ to $\Dir(\E)$.
On the other hand, the map $h_1\colon\Ell(\E)\to\Ell(\E)$ is homotopic to $\Id_{\Ell(\E)}$ via the homotopy $h_s$.
It follows that $h_1\colon\Ell(\E)\to\Dir(\E)$ 
is homotopy inverse to the embedding $\Dir(\E)\hookto\Ell(\E)$,
that is, this embedding is a bundle homotopy equivalence.
This completes the proof of the proposition.
\endproof

\begin{prop}\label{prop:retr3}
For every $\E\in\VectXM$ 
the natural embeddings $\Gam\Dir(\E)\hookto\Gam\Ell(\E)$ and $\Gam\Dirt(\E)\hookto\Gam\Ellt(\E)$
are homotopy equivalences.
Moreover, 
\begin{enumerate}\upskip
	\item There exists a deformation retraction of $\Gam\Ell(\E)$ onto a subspace of $\Gam\Dir(\E)$ preserving $\EEm(\gamma)$.
	\item There exists a deformation retraction of $\Gam\Ellt(\E)$ onto a subspace of $\Gam\Dirt(\E)$
preserving both $\EEm(\gamma)$ and $\F(\gamma)$.
\end{enumerate}
\end{prop}

\proof
1. The fiberwise deformation retraction $h$ from Proposition \ref{prop:retr2} 
induces the deformation retraction $H$ on the space of sections
satisfying the conditions of the proposition.

2. Denote by $p$ the natural projection $\Gam\Ellt(\E)\to\Gam\Ell(\E)$, which forgets boundary conditions.
We define the deformation retraction $\bar{H}\colon[0,1]\times\Gam\Ellt(\E)\to\Gam\Ellt(\E)$
by the formula $\bar{H}_s(\gamma)(x)=(H_s(p\gamma)(x),T(x))$ for $\gamma\colon x\mapsto(A(x),T(x))$.
Since $\EEm(H_s(p\gamma)) = \EEm(\gamma)$, $\bar{H}_s(\gamma)$ is well defined.
By definition of $\bar{H}$, the subbundles $\F(\bar{H}_s(\gamma))$ and $\F(\gamma)$ of $\EEmY(\gamma)$
coincide for every $s\in[0,1]$ and $\gamma\in\Gam\Ellt(\E)$.

3. The fiberwise homotopy $h'$ from Proposition \ref{prop:retr2}
induces the homotopy between the restriction of $H_1$ to $\Gam\Dir(\E)$ and the identity map of $\Gam\Dir(\E)$,
as well as the homotopy between the restriction of $\bar{H}_1$ to $\Gam\Dirt(\E)$ and the identity map of $\Gam\Dirt(\E)$.
The same arguments as in the proof of Proposition \ref{prop:retr2}
show that $H_1\colon\Gam\Ell(\E)\to\Gam\Dir(\E)$ is homotopy inverse 
to the embedding $\Gam\Dir(\E)\hookto\Gam\Ell(\E)$
and $\bar{H}_1\colon\Gam\Ellt(\E)\to\Gam\Dirt(\E)$ is homotopy inverse 
to the embedding $\Gam\Dirt(\E)\hookto\Gam\Ellt(\E)$.
This completes the proof of the proposition. 
\endproof

\sub{Retraction of special subspaces.}
The following proposition is one of the key ingredients in the proof of the index theorem.

\begin{prop}\label{prop:def-pm}
There exists a deformation retraction of $\Gam\Ellp(\E)$ onto a subspace of $\Gam\Dirp(\E)$
and a deformation retraction of $\Gam\Ellm(\E)$ onto a subspace of $\Gam\Dirm(\E)$.
\end{prop}

\proof
Let $\bar{H}$ be a deformation retraction of $\Gam\Ellt(\E)$ onto a subspace of $\Gam\Dirt(\E)$
satisfying the conditions of Proposition \ref{prop:retr3}.
For $\gamma\in\Gam\Ellp(\E)$ and $\gamma_s = \bar{H}_s(\gamma)$ we have $\F(\gamma_s)=\F(\gamma) = 0$, 
so by Proposition \ref{prop:Fmp} $\gamma_s\in\Gam\Ellp(\E)$ for all $s$.
In particular, $\gamma_1\in\Gam\Ellp(\E)\cap\Gam\Dirt(\E) = \Gam\Dirp(\E)$.
For $\gamma\in\Gam\Ellm(\E)$ and $\gamma_s = \bar{H}_s(\gamma)$ we have 
$\F(\gamma_s) = \F(\gamma) = \EEm(\gamma)=\EEm(\gamma_s)$,
so by Proposition \ref{prop:Fmp} $\gamma_s\in\Gam\Ellm(\E)$ for all $s$.
In particular, $\gamma_1\in\Gam\Ellm(\E)\cap\Gam\Dirt(\E) = \Gam\Dirm(\E)$.
This completes the proof of the proposition.
\endproof

\section{Index theorem}\label{sec:ind-proof}

\upskip

\sub{Invertible Dirac operators.}
We have no means to detect the invertibility of an arbitrary element of $\Ellt(E)$ by purely topological methods.
However, there is a big class of \textit{odd Dirac} operators which are necessarily invertible:

\begin{prop}[\cite{Pr17}, Proposition 10.1]\label{prop:zero1}
If $(A,L)$ is an element of $\Dirp(E)$ or $\Dirm(E)$, then $A_L$ has no zero eigenvalues.
In other words, both $\Dirp(E)$ and $\Dirm(E)$ are subspaces of $\Ellt^0(E)$.
\end{prop}

\upskip
\sub{Vanishing of the analytical index.}
Taking into account Proposition \ref{prop:def-pm}, 
we are now able to describe, in purely topological terms, a big class of sections of $\Ellt(\E)$
which are homotopic to families of invertible operators.

\begin{prop}\label{prop:ind0}
Let $\gamma$ be an element of $\Gammapm\Ellt(\E)$ or $\Gammabox\Ellt(\E)$.
Then $\gamma$ is homotopic to a section of $\Ellt^0(\E)$, 
and hence $\inda(\gamma) = 0$.
\end{prop}

\proof
1. If $\gamma\in\Gammapm\Ellt(\E)$,
then $\gamma = \gamma'\oplus\gamma''$
with $\gamma'\in\Gam\Ellp(\E')$ and $\gamma''\in\Gam\Ellm(\E'')$
for some orthogonal decomposition $\E\cong\E'\oplus\E''$.
By Proposition \ref{prop:def-pm}, $\gamma'$ is homotopic to some $\gamma'_1\in\Gam\Dirp(\E')$
and $\gamma''$ is homotopic to some $\gamma''_1\in\Gam\Dirm(\E'')$.
By Proposition \ref{prop:zero1}, $\gamma'_1$ and $\gamma''_1$ 
are sections of $\Ellt^0(\E')$ and $\Ellt^0(\E'')$ respectively.
It follows that $\gamma$ is homotopic to $\gamma'_1\oplus\gamma''_1$, which is a section of $\Ellt^0(\E)$.

2. Suppose that $\gamma = 1_W\boxtimes(A,L)$ for some $(A,L)\in\Ellt(E)$ and $W\in\VectX$.
Since $A_L$ is Fredholm, $A_L-\lambda$ is invertible for some $\lambda\in\R$,
that is, $(A-\lambda,L)\in\Ellt^0(E)$.
The path $h\colon[0,1]\to\Gammabox\Ellt(W\boxtimes E)$ 
given by the formula $h_s = 1_W\boxtimes(A-s\lambda,L)$
connects $\gamma$ with $1_W\boxtimes(A-\lambda,L)\in\Ellt^0(W\boxtimes E)$.

In the general case, for $\gamma = \bigoplus{1_{W_i}\boxtimes(A_i,L_i)}\in\Gammabox\Ellt(\E)$, 
we take such a homotopy as described above for every direct summand 
$1_{W_i}\boxtimes(A_i,L_i)$ independently.
The direct sum of these homotopies gives a required homotopy of $\gamma$ 
to a section $h_1(\gamma)\in\Gammabox\Ellt^0(\E)$. 
	
3. It follows from the homotopy invariance of the analytical index and its vanishing on sections of $\Ellt^0(\E)$
that $\inda(\gamma) = 0$.
\endproof

\sub{Index theorem.}
Now we are able to prove our index theorem.

\begin{thm}\label{thm:ind}
Let $X$ be a compact space and $\E\in\VectXM$. 
Then the analytical index is equal to the topological index for every section $\gamma$ of $\Ellt(\E)$:
\begin{equation}\label{eq:thm1}
\inda(\gamma) = \indt(\gamma).
\end{equation}
In particular, this equality holds for every continuous map $\gamma\colon X\to\Ellt(E)$, $E\in\VectM$.
\end{thm}

\proof
By Proposition \ref{prop:prop-inda} $\Phi=\inda$ satisfies conditions (E0--E4). 
By Proposition \ref{prop:ind0} $\Phi$ satisfies conditions (\Epm, \Ebox).
By Theorem \ref{thm:top-fam2} there is an integer $m$ such that $\inda(\gamma) = m\cdot\indt(\gamma)$
for every section $\gamma$ of $\Ellt(\E)$, every $\E\in\VectXM$, and every compact space $X$.
The factor $m$ does not depend on $X$, but can depend on $M$.

For $X=S^1$ the analytical index of $\gamma$ coincides with the spectral flow $\spf(\gamma)$ 
by Proposition \ref{prop:prop-inda},
while the topological index of $\gamma$  coincides with $c_1(\F(\gamma))[\pMS]$ by Proposition \ref{prop:prop-indt}.
Hence it is sufficient to compute the quotient 
\[ m = m(M) = \frac{\spf(\gamma)}{c_1(\F(\gamma))[\pMS]} \]
for some loop $\gamma\colon S^1\to\Dirt(2k_M)$ such that the denominator of this quotient does not vanish.

For the case of an annulus this computation was performed by the author in \cite[Theorem 4]{Pr11} by direct evaluation;
it was shown there that the factor $m$ for the annulus is equal to 1.
Moreover,  the value of $m(M)$ is the same for all surfaces $M$ \cite[Lemmas 11.3 and 11.5]{Pr17}.
These two results together imply that $m(M)=1$ for any surface $M$.
Therefore, $\inda(\gamma) = \indt(\gamma)$, which completes the proof of the theorem.
\endproof

\section{Universality of the analytical index}\label{sec:uni}

Recall that we denoted by $\Ellt^0(E)$ the subspace of $\Ellt(E)$ consisting of all pairs $(A,L)$
such that the unbounded operator $A_L$ has no zero eigenvalues,
and by $\Ellt^0(\E)$ the subbundle of $\Ellt(\E)$ whose fiber over $x\in X$ is $\Ellt^0(\E_x)$.
Sections of $\Ellt^0(\E)$ correspond to families of invertible self-adjoint elliptic boundary problems.

\begin{thm}\label{thm:an-fam}
  Let $X$ be a compact space,
	and let $\gamma_i$ be a section of $\Ellt(\E_i)$, $\E_i\in\VectXM$, $i=1,2$.
	Then the following two conditions are equivalent:
\begin{enumerate}\upskip
	\item $\inda(\gamma_1) = \inda(\gamma_2)$.
	\item There are $k\in\N$, sections $\beta^0_i$ of $\Ellt^0(2\kXM)$, 
	and sections $\gamma^0_i$ of $\Ellt^0(\E_i)$ such that 
	$\gamma_1\oplus\gamma^0_2\oplus\beta^0_1$ and 
	$\gamma^0_1\oplus\gamma_2\oplus\beta^0_2$ 
	are homotopic sections of $\Ellt(\E_1\oplus\E_2\oplus 2\kXM)$.
\end{enumerate}
\end{thm}

\proof
($2\Rightarrow 1$) follows immediately from properties (I0--I2) of the family index.

Let us prove ($1\Rightarrow 2$).
By Theorem \ref{thm:ind} the equality $\inda(\gamma_1) = \inda(\gamma_2)$ 
implies $\indt(\gamma_1) = \indt(\gamma_2)$.
By Theorem \ref{thm:uni-top} there are 
$\beta_i^{\pm}\in\Gammapm\Dirt(2n_{X,M})$ and $\beta_i^{\boxtimes}\in\Gammabox\Dirt(2l_{X,M})$, $i=1,2$, 
such that the direct sums 
$\gamma_1\oplus\gamma_2^+\oplus\beta_1^{\pm}\oplus\beta_1^{\boxtimes}$ and
$\gamma_1^+\oplus\gamma_2\oplus\beta_2^{\pm}\oplus\beta_2^{\boxtimes}$
are homotopic.
By Proposition \ref{prop:ind0} 
$\gamma_i^+$ is homotopic to a section $\gamma^0_i$ of $\Ellt^0(\E_i)$
and $\beta_i^{\pm}\oplus\beta_i^{\boxtimes}$ is homotopic to a section 
$\beta^0_i$ of $\Ellt^0(2k_{X,M})$, $k=n+l$.
This completes the proof of the theorem.
\endproof

\sub{Universality for families.}
In Section \ref{sec:uni-top} we considered invariants satisfying properties (\Epm, \Ebox) and (E1-E4).
Now we replace the topological properties (\Epm, \Ebox) by the following analytical property:

(E0) $\Phi$ vanishes on sections of $\Ellt^0(\E)$.

\begin{thm}\label{thm:uni-fam}
Let $X$ be a compact topological space and $\Lambda$ be a commutative monoid.
Let $\Ve$ be a subclass of $\VectXM$ satisfying condition \eqref{eq:Ve}.
Suppose that we associate an element $\Phi(\gamma)\in\Lambda$
with every section $\gamma$ of $\Ellt(\E)$ for every $\E\in\Ve$.
Then the following two conditions are equivalent:
\begin{enumerate}\upskip
	\item $\Phi$ satisfies properties (E0--E2). 
	\item $\Phi$ has the form $\Phi(\gamma) = \vartheta(\inda(\gamma))$
        for some (unique) monoid homomorphism $\vartheta\colon K^1(X)\to\Lambda$.
\end{enumerate}\upskip
\end{thm}

\proof
($2\Rightarrow 1$) follows from properties (I0--I2) of the family index.
($1\Rightarrow 2$) follows from Theorem \ref{thm:uni-top}, Proposition \ref{prop:ind0}, and Theorem \ref{thm:ind}.
\endproof

\begin{thm}\label{thm:uni-fam2}
Suppose that we associate an element $\Phi_X(\gamma)\in K^1(X)$
with every section $\gamma$ of $\Ellt(\E)$ 
for every compact space $X$ and every $\E\in\VectXM$.
Then the following two conditions are equivalent:
\begin{enumerate}\upskip
	\item The family $\Phi = (\Phi_X)$ satisfies properties (E0--E4).				
	\item $\Phi$ has the form $\Phi_X(\gamma) = m\cdot\inda(\gamma)$ for some integer $m$.
\end{enumerate}\upskip
\end{thm}

\proof
($2\Rightarrow 1$) follows from properties (I0--I4) of the family index.
($1\Rightarrow 2$) follows from Theorem \ref{thm:top-fam2}, Proposition \ref{prop:ind0}, and Theorem \ref{thm:ind}.
\endproof

\sub{Universality for maps.}
Theorem \ref{thm:an-fam} applied to trivial bundles $\E_1$ and $\E_2$ takes the following form.

\begin{thm}\label{thm:uni-maps-n}
  Let $X$ be a compact space
	and $\gamma\colon X\to\Ellt(2k_M)$, $\gamma'\colon X\to\Ellt(2k'_M)$ be continuous maps.
	Then the following two conditions are equivalent:
\begin{enumerate}\upskip
	\item $\inda(\gamma) = \inda(\gamma')$.
	\item There are $n\in\N$ and maps $\beta\colon X\to\Ellt^0(2(n-k)_M)$, $\beta'\colon X\to\Ellt^0(2(n-k')_M)$ 
	such that 
	the maps $\gamma\oplus\beta$ and $\gamma'\oplus\beta'$ from $X$ to $\Ellt(2n_M)$ are homotopic.
\end{enumerate}
\end{thm}

Theorem \ref{thm:uni-fam} applied to the set $\Ve=\set{2\kXM}$ of trivial bundles takes the following form. 

\begin{thm}\label{thm:uni-maps1}
Let $X$ be a compact space and $\Lambda$ be a commutative monoid.
Suppose that we associate an element $\Phi(\gamma)\in\Lambda$
with every map $\gamma\colon X\to\Ellt(2k_M)$ for every integer $k$.
Then the following two conditions are equivalent:
\begin{enumerate}\upskip
	\item $\Phi$ is homotopy invariant, additive with respect to direct sums, and vanishes on maps to $\Ellt^0(2k_M)$.
	\item $\Phi$ has the form $\Phi(\gamma) = \vartheta(\inda(\gamma))$
        for some (unique) monoid homomorphism $\vartheta\colon K^1(X)\to\Lambda$.
\end{enumerate}
\end{thm}

\upskip

\sub{The analytical index as a homomorphism.}
Denote by $\Ellt^0_M$ the disjoint union of subspaces $\Ellt^0(2k_M)\subset\Ellt(2k_M)$ for all $k\in\N$;
it is a subsemigroup of $\Ellt_M$.
The inclusion $\Ellt^0_M\subset\Ellt_M$ induces the homomorphism $[X,\Ellt^0_M]\to[X,\Ellt_M]$;
we denote by $[X,\Ellt_M]^0$ its image.

Since the analytical index is additive with respect to direct sums, 
it defines the monoid homomorphism $\inda\colon C(X,\Ellt_M)\to K^1(X)$.
Since the analytical index is homotopy invariant, 
this homomorphism factors through the homomorphism $C(X,\Ellt_M)\to[X;\Ellt_M]$.
Since the analytical index vanishes on maps to $\Ellt_M^0$, 
it factors through $[X,\Ellt_M]/[X,\Ellt_M]^0$.
In other words, there exists a monoid homomorphism 
$$\kappa_a\colon [X,\Ellt_M]/[X,\Ellt_M]^0 \to K^1(X)$$ 
such that the following diagram is commutative:
\begin{equation}\label{diag:kappa}							
	\begin{tikzcd}
		C(X,\Ellt_M) \arrow{r} \arrow[swap]{drr}{\inda} &
		{[X,\Ellt_M]}  \arrow{r} & 
		{[X,\Ellt_M]/[X,\Ellt_M]^0} \arrow{d}{\kappa_a} 
		\\
		 &  &  K^1(X) 
	\end{tikzcd}
\end{equation}

\begin{thm}\label{thm:uni-maps2}
Let $X$ be a compact space.
Then $[X,\Ellt_M]/[X,\Ellt_M]^0$ is an Abelian group isomorphic to $K^1(X)$,
and the homomorphism $\kappa_a$ on diagram \eqref{diag:kappa} is an isomorphism.
\end{thm}

\proof
Denote the commutative monoid $[X,\Ellt_M]/[X,\Ellt_M]^0$ by $\Lambda$
and the composition of horizontal arrows on diagram \eqref{diag:kappa} by $\Phi$. 
The homomorphism $\Phi$ factor through $[X,\Ellt_M]$ and vanishes on maps to $\Ellt^0(2k_M)$.
By Theorem \ref{thm:uni-maps1} $\Phi = \vartheta\circ\inda$
for some (unique) monoid homomorphism $\vartheta\colon K^1(X)\to\Lambda$.
By definition, $\Phi$ is surjective.
By Theorem \ref{thm:ind} $\inda=\indt$; by Proposition \ref{prop:prop-indt}(T5) $\indt$ is surjective.
Thus $\kappa_a$ and $\vartheta$ are mutually inverse monoid homomorphisms,
so $\kappa_a$ is an isomorphism. 
This completes the proof of the theorem.
\endproof

\appendix{} 

\section*{Appendices}

\addtocontents{toc}{\vspace{20pt}Appendices\vspace{-18pt} \par}

\section{Smoothing}\label{sec:smooth}

This appendix is devoted to the proof of two technical results,
Propositions \ref{prop:Vany} and \ref{prop:Vhom},
that are used in the main part of the paper.

\sub{Smoothing of maps.}
Let $Z$ and $Z'$ be compact smooth manifolds and $r$ be a non-negative integer.
We denote by $C^{r,\infty}(Z,Z')$ the space $\Cinf(Z,Z')$ of smooth maps from $Z$ to $Z'$
equipped with the topology induced by the natural inclusion $\Cinf(Z,Z')\hookto C^r(Z,Z')$.

\begin{prop}\label{prop:smooth}
Let $X$ be a compact space and $Z$, $Z'$ be compact smooth manifolds. 
Then for every non-negative integer $r$ the following statements hold: 
\begin{enumerate}\upskip
	\item The space $C^{r,\infty}(Z,Z')$ is locally contractible. 
	\item The space $C(X\times Z, Z') = C(X,C(Z,Z'))$ is locally contractible and 
		    contains $C(X,\Cinf(Z,Z'))$ as a dense subset.
				In particular, every $f\in C(X,C(Z,Z'))$ is homotopic to some $F\in C(X,\Cinf(Z,Z'))$.
	\item If continuous maps $f_0,f_1\colon X\to \Crinf(Z,Z')$ are homotopic as maps from $X$ to $C(Z,Z')$,
	      then they are homotopic as maps from $X$ to $C^{r,\infty}(Z,Z')$.
				Moreover, $\H^r(f_0,f_1)$ is a dense subset of $\H^0(f_0,f_1)$, 
				where $\H^r(f_0,f_1)$ denotes the subspace of $C([0,1]\times X, \Crinf(Z,Z'))$ 
				consisting of maps $f$ such that $\restr{f}{\set{i}\times X}=f_i$ for $i=0,1$.
\end{enumerate}
\end{prop}

\proof
Let us choose a smooth embedding of $Z'$ to $\R^n$ for some $n$;
let $p\colon N\to Z'$ be its normal bundle. 
Denote by $N_{\varepsilon}$ the $\varepsilon$-neighborhood of the zero section in $N$.

Let $\varepsilon>0$ be small enough, so that the restriction of the geodesic map $q\colon N\to\R^n$
to $N_{\varepsilon}$ is an embedding.
This embedding allows to identify $N_{\varepsilon}$ with the $\varepsilon$-neighborhood of $Z'$ in $\R^n$.
We denote the restriction of $p$ to $N_{\varepsilon}$ again by $p$;
we will use only this small part of the normal bundle from now on.
The map $p$ takes a point $u\in N_{\varepsilon}$ to the (unique) closest point on $Z'$.

\textbf{2a.} Let $f$ be an arbitrary element of $C(X\times Z, Z')$.
For every $s\in[0,1]$ and every two points $u,v\in Z'$ such that $\norm{u-v}_{\R^n} < \varepsilon$,
the point $w = su+(1-s)v$ lies in $N_{\varepsilon}$,
$\norm{w-p(w)} = d(w,Z')\leq\norm{w-v}$, and
$$\norm{p(w)-u} = \norm{p(w)-w+w-u} \leq \norm{v-w}+\norm{w-u} = \norm{v-u} < \varepsilon,$$
so $p(w)$ lies in the $\varepsilon$-neighborhood of $u$.
Thus the formula 
\begin{equation}\label{eq:hsg}
	h^i_s(g) = p\circ(sf+(1-s)g)
\end{equation}
defines the contracting homotopy of the $\varepsilon$-neighborhood 
$$U_{f,\,\varepsilon} = \set{ g\in C(X\times Z, Z')\colon \norm{g-f}_{C(X\times Z,\,\R^n)} < \varepsilon }$$ 
of $f$ in $C(X\times Z, Z')$.
It follows that $C(X\times Z, Z')$ is locally contractible.

\textbf{1.} If $f\in C(X,\Crinf(Z,Z'))$, then formula \eqref{eq:hsg} defines the contracting homotopy 
of $C(X,\Crinf(Z,Z')) \cap U_{f,\,\varepsilon}$ to $f$. 
In the particular case of a one-point space $X$ this implies the first claim of the proposition.

\textbf{2b.} For every $y\in X$ choose $g_y\in\Cinf(Z,\R^n)$ such that 
$\norm{g_y-f(y)}_{C(Z,\,\R^n)} < \varepsilon$. 
Then 
\begin{equation}\label{eq:Xy}
	X_y = \set{ x\in X\colon \norm{g_y-f(x)}_{C(Z,\,\R^n)} < \varepsilon }
\end{equation}
is an open neighborhood of $y$.
Since $X$ is compact, the open covering $(X_y)_{y\in X}$ of $X$ contains a finite sub-covering $(X_y)_{y\in I}$.
Choose a partition of unity $(\rho_y)_{y\in I}$ subordinated to this finite covering.
We define the map $g'\colon X\to \Cinf(Z,\R^n)$ by the formula
$g'(x) = \sum_{y\in I}\rho_y(x)g_y$.
Obviously, $g'$ is continuous.
By \eqref{eq:Xy}, $\norm{g'(x)(z)-f(x)(z)} < \varepsilon$ for every $x\in X$, $z\in Z$, 
so the image of $g'$ lies in $\Cinf(Z,N_{\varepsilon})$.
The composition $g = p\circ g'$ is a continuous map from $X$ to $\Cinf(Z,Z')$. 
Moreover, $g$ and $f$ are homotopic as continuous maps from $X\times Z$ to $Z'$,
with a homotopy given by the formula \eqref{eq:hsg}. 
This proves the density of $C(X,\Cinf(Z,Z'))$ in $C(X,C(Z,Z'))$
and completes the proof of the second claim of the proposition.

\textbf{3.} Let $f\colon [0,1]\times X\to C(Z,Z')$ be a homotopy between $f_0,f_1\in C(X,\Crinf(Z,Z'))$.
By the second claim of the proposition, $C([0,1]\times X,\Cinf(Z,Z'))$ is dense in $C([0,1]\times X,C(Z,Z'))$.
Thus there is a continuous map $F\colon [0,1]\times X\to \Crinf(Z,Z')$ such that
$\norm{F-f}_{C([0,1]\times X\times Z,\,\R^n)} < \varepsilon$.
The last inequality implies $\norm{F_i-f_i}_{C(X\times Z,\,\R^n)} < \varepsilon$ for $i=0,1$,
where $F_i = \restr{F}{\set{i}\times X}$.
Applying again the second claim of the proposition, 
we obtain a homotopy $h^{(i)}\colon [0,1]\times X\to \Crinf(Z,Z')$ between $F_i$ and $f_i$ such that 
$\norm{h^{(i)}_s-f_i}_{C(X\times Z,\,\R^n)} < \varepsilon$ for all $s\in[0,1]$.
Concatenating $h^{(0)}$, $F$, and $h^{(1)}$ and suitably reparametrizing the result,
we obtain the path in $C(X,C^{r,\infty}(Z,Z'))$ connecting $f_0$ with $f_1$ 
and lying in the $\varepsilon$-neighborhood of $f$.
This proves the third claim of the proposition.
\endproof

\sub{Smoothing of subbundles.}
Let us recall some designations from the main part of the paper.
Let $X$ be a topological space and $Z$ be a smooth manifold.
We denoted by $\Vect_{\,X,\,Z}$ the class of all locally trivial fiber bundles $\E$ over $X$,
whose fiber $\E_x$ is a smooth Hermitian vector bundle over $Z$ for every $x\in X$ 
and the structure group is the group $U(\E_x)$ of smooth unitary bundle automorphisms of $\E_x$ 
equipped with the $C^1$-topology.
We say that $\W\subset \V$ is a subbundle of $\V\in\Vect_{\,X,\,Z}$ 
if $\W\in\Vect_{\,X,\,Z}$ and $\W_x$ is a smooth subbundle of $\V_x$ for every $x\in X$.
For $\V\in\Vect_{\,X,\,Z}$ we denoted by $\bra{\V}$ the vector bundle over $X\times Z$
whose restriction to $\set{x}\times Z$ is the fiber $\V_x$ with the forgotten smooth structure.
Similarly, for a subbundle $\W$ of $\V$ we denote by $\bra{\W}$ the corresponding vector subbundle of $\bra{\V}$.

\begin{prop}\label{prop:Vany}
Let $X$ be a compact space, $Z$ be a compact smooth manifold,
and $V$ be a subbundle of a trivial vector bundle $k_{X\times Z}$.
Then $V$ is homotopic to $\bra{\V}$ for some subbundle $\V$ of $k_{\,X,\,Z}$.
In particular, every vector bundle over $X\times Z$ is isomorphic to $\bra{\V}$ 
for some $\V\in\Vect_{\,X,\,Z}$.
\end{prop}

\proof
Let $f\colon X\times Z\to\Gr(\CC^k)$ be the continuous map corresponding to the embedding $V\hookto k_{X\times Z}$.
By Proposition \ref{prop:smooth}(2), $f$ considered as a map from $X$ to $C(Z,\Gr(\CC^k))$
is homotopic to a continuous map $F\colon X\to C^{1,\infty}(Z,\Gr(\CC^k))$.
Such a map $F$ defines a fiber bundle $\V$ over $X$, 
whose fiber $\V_x$ is a smooth subbundle of $k_{Z}$ 
given by the smooth map $F(x)\colon Z\to\Gr(\CC^k)$.
A homotopy between $F$ and $f$ induces the homotopy between the vector subbundles $\bra{\V}$ and $V$ of $k_{X\times Z}$.

Let $x_0$ be an arbitrary point of $X$ and $F_0=F(x_0)$. 
By Proposition \ref{prop:smooth}(1), 
there is a contractible neighbourhood $U'$ of $F_0$ in $C^{1,\infty}(Z,\Gr(\CC^k))$.
Let $h$ be a corresponding contracting homotopy. 
Then the restriction of $F$ to $U=F\inv(U')\subset X$ is homotopic, 
as a map from $U$ to $C^{1,\infty}(Z,\Gr(\CC^k))$, 
to the constant map $U\ni x\mapsto F_0$, with the homotopy $H_s(x) = h_s(F(x))$. 
It follows that the restriction of $\V$ to $U$ is a trivial bundle. 
Thus $\V\in\Vect_{\,X,\,Z}$ and $\V$ is a subbundle of $k_{\,X,\,Z}$,
which completes the proof of the proposition.
\endproof

\begin{prop}\label{prop:Vhom}
Let $X$ and $Z$ be as in Proposition \ref{prop:Vany}. 
Let $\E\in\Vect_{\,X,\,Z}$ and $\V_0$, $\V_1$ be subbundles of $\E$.
Suppose that $\bra{\V_0}$ and $\bra{\V_1}$ are homotopic as subbundles of $\bra{\E}$.
Then $\V_0$ and $\V_1$ are homotopic subbundles of $\E$.
\end{prop}

\proof
Consider first the case of a trivial $\E = k_{\,X,\,Z}$.
Then $\V_i$ can be identified with a continuous map $F_i\colon X\to\Crinf(Z,\Gr(\CC^k))$, $i=1,2$.
Since $\bra{\V_0}$ and $\bra{\V_1}$ are homotopic as subbundles of $\bra{\E}$,
$F_0$ and $F_1$ are homotopic as maps from $X$ to $C(Z,\Gr(\CC^k))$.
By Proposition \ref{prop:smooth}(3), they are homotopic as maps from $X$ to $\Crinf(Z,\Gr(\CC^k))$.
It follows that $\V_0$ and $\V_1$ are homotopic subbundles of $\E$.

Let now $\E$ be an arbitrary element of $\Vect_{\,X,\,Z}$.

Denote by $\Gat\E$ the vector space of continuous maps $X\ni x\mapsto \Gamma^{1,\infty}\E_x$,
where $\Gamma^{1,\infty}\E_x$ denotes the space of smooth sections of $\E_x$ with the $C^1$-topology.
It is finitely generated as an $\A$-module, where $\A = C(X,C^{1,\infty}(Z,\CC))$.
Indeed, let $(X_i)$ be a finite open covering of $X$ such that 
the restriction $\E_i$ of $\E$ to $X_i$ is a trivial bundle with a fiber $E_i$.
Let $(\rho_i)$ be a partition of unity subordinated to this finite covering,
and let $(v_{ij})$ be a finite generating set for $\Ginf E_i$.	
Then $u_{ij} = \rho_i v_{ij}$ form a finite generating set for $\Gat\E$.

Let $(u_i)_{i=1}^k$ be a finite generating set for the $\A$-module $\Gat\E$.
For every $x\in X$, the set $(u_i(x))$ of smooth sections of $\E_x$ 
generates $\Ginf\E_x$ as a $\Cinf(Z,\CC)$-module and thus defines 
the smooth surjective bundle morphism $\pi_x\colon k_Z\to\E_x$ continuously depending on $x$.
Then the kernel $\K_x$ of $\pi_x$ continuously depends on $x$ and is locally trivial. 
Thus the family $(\K_x)$ of smooth vector subbundles of $k_Z$ defines the subbundle $\K$ of $k_{\,X,\,Z}$.
Denote by $K$ the continuous map from $X$ to $C^{1,\infty}(Z,\Gr(\CC^k))$ corresponding to $\K$.
Obviously, subbundles of $\E$ are in one-to-one correspondence with subbundles of $k_{\,X,\,Z}$ containing $\K$.

Let $\V_0$, $\V_1$ be subbundles of $\E$.
Denote by $\W_0$, $\W_1$ the corresponding subbundles of $k_{\,X,\,Z}$
and by $F_0,F_1$ the corresponding maps from $X$ to $C^{1,\infty}(Z,\Gr(\CC^k))$.
If $\bra{\V_0}$ and $\bra{\V_1}$ are homotopic as subbundles of $\bra{\E}$,
then there is a homotopy $h\colon[0,1]\times X\to C(Z,\Gr(\CC^k))$ between $F_0$ and $F_1$ such that 
$h_s(x)(z)\supset K(x)(z)$ for every $s\in[0,1]$, $x\in X$, and $z\in Z$.

Equip $\Gr(\CC^k)$ with a smooth Riemannian metric.
For $L\in\Gr(\CC^k)$ denote by $\Gr_L(\CC^k)$ the submanifold of $\Gr(\CC^k)$
consisting of subspaces of $\CC^k$ containing $L$.
Denote by $p_L\colon N_L\to\Gr_L(\CC^k)$ the normal bundle of $\Gr_L(\CC^k)$ in $\Gr(\CC^k)$, 
and by $N_{L,\varepsilon}$ the $\varepsilon$-neighborhood of the zero section in $N_L$.
Let $\varepsilon>0$ be small enough, so that for every $L\in\Gr(\CC^k)$
the geodesic map $q_L\colon N_{L,\varepsilon}\to\Gr(\CC^k)$ is an embedding.
Similarly to the proof of Proposition \ref{prop:smooth},
we identify $N_{L,\varepsilon}$ with the $\varepsilon$-neighborhood of $\Gr_L(\CC^k)$ in $\Gr(\CC^k)$.
The map $p_L$ smoothly depends on $L$ with respect to this identification.

By Proposition \ref{prop:smooth}(3), there is a homotopy $H\colon[0,1]\times X\to C^{1,\infty}(Z,\Gr(\CC^k))$ 
between $F_0$ and $F_1$ such that the distance between $H_s(x)(z)$ and $h_s(x)(z)$ 
is less then $\varepsilon$ for all $s$, $x$, and $z$.
Then the continuous map $F\colon[0,1]\times X\to C^{1,\infty}(Z,\Gr(\CC^k))$ 
defined by the formula $F_s(x)(z) = p_{K(x)(z)}(H_s(x)(z))$ 
is a homotopy between $F_0$ and $F_1$ such that 
$F_s(x)(z)\supset K(x)(z)$ for every $s$, $x$, and $z$.
Thus $F$ defines the homotopy $(\W_s)$ between $\W_0$ and $\W_1$ 
such that $\K$ is a subbundle of $\W_s$ for every $s\in[0,1]$.
Factoring by $\K$, we obtain the homotopy $(\V_s)$ between $\V_0$ and $\V_1$ as subbundles of $\E$,
which completes the proof of the proposition.
\endproof

\section{Natural transformations of $K^1$}\label{sec:K1}

\renewcommand{\thesection}{\Alph{section}} 

The purpose of this Appendix is the proof of the following result, which we use in the main part of the paper.

Let $\Cat$ be one of the following categories:
the category of compact Hausdorff spaces and continuous maps,
the category of finite CW-complexes and continuous maps, 
or the category of smooth closed manifolds and smooth maps. 
We consider $K^1$ as a functor from $\Cat$
to the category of Abelian groups.

\begin{prop}\label{prop:K1}
  Let $\vartheta$ be a natural self-transformation of the functor $X\mapsto K^1(X)$
	respecting the $K^0(\cdot)$-module structure
	(that is, $\vartheta(\lambda\mu) = \lambda\vartheta(\mu)$ 
	for every object $X$ of $\Cat$ and every $\lambda\in K^0(X)$, $\mu\in K^1(X)$).
	Then $\vartheta$ is multiplication by some integer $m$, that is,
	$\vartheta_X(\mu)=m\mu$ for every object $X$ of $\Cat$ and every $\mu\in K^1(X)$.
	In particular, if $\vartheta_{S^1}$ is the identity, then $\vartheta_X$ is the identity for every $X$.
\end{prop}

\proof
$K^1(U(1))$ is an infinite cyclic group, 
so $\vartheta_{U(1)}$ is multiplication by some integer; denote this integer by $m$.

Let $X$ be an object of $\Cat$ and $\mu\in K^1(X)$. 
There is $n\in N$ and a continuous map $f\colon X\to U(n)$ such that $\mu=f^*\beta$,
where $\beta$ denotes the element of $K^1(U(n))$ corresponding to the canonical representation $U(n)\to\Aut(\CC^n)$.
If $X$ is a smooth manifold, then $f$ can be chosen to be smooth.
Since $\vartheta$ is natural, $\vartheta_X\mu = f^*(\vartheta_{U(n)}\beta)$.
Therefore, it is sufficient to show that $\vartheta_{U(n)}\beta = m\beta$. 

Let $T=U(1)^n$ be the maximal torus in $U(n)$ consisting of diagonal matrices 
and $V = U(n)/T$ be the flag manifold.
Let $\pi\colon V\times T \to U(n)$ be the natural projection given by the formula
$\pi(gT,u) = gug\inv$.

Denote by $L_1,\ldots,L_n$ the canonical linear bundles over $V$,
and let $l_i=[L_i]\in K^0(V)$.
Let $\alpha_i$ be the element of $K^1(T)$ corresponding to the projection of $T=U(1)^n$ on the $i$-th factor.
We denote the liftings of $L_i$, $l_i$, and $\alpha_i$ to $V\times T$ by the same letters.
The lifting of $\beta$ can be written in these notations as $\pi^*\beta = \sum_{i=1}^n{l_i\alpha_i}$.

The element $\alpha_i$ is lifted from $U(1)$ and $\vartheta_{U(1)}$ is multiplication by $m$,
hence $\vartheta_{V\times T}(\alpha_i) = m\alpha_i$. 
Since $\vartheta_{V\times T}$ is a $K^0(V\times T)$-module homomorphism, we have
\begin{equation*}
  \pi^*(\vartheta_{U(n)}\beta) = \vartheta_{V\times T}(\pi^*\beta) = 
   \sum_{i=1}^n \vartheta_{V\times T}\br{{l_i\alpha_i}} = 
	 \sum_{i=1}^n{l_i\cdot\vartheta_{V\times T}(\alpha_i)} = 
	 \sum_{i=1}^n{l_i\cdot m\alpha_i} = \pi^*(m\beta),	
\end{equation*}
that is, $\pi^*\br{\vartheta_{U(n)}\beta-m\beta}=0$.
To complete the proof of the proposition, 
it is sufficient to show the injectivity of the homomorphism $\pi^* \colon K^1(U(n)) \to K^1(V\times T)$,
which we perform in the following lemma.

\begin{lem}\label{lem:inj}
The homomorphism $\pi^* \colon K^*(U(n)) \to K^*(V\times T)$ is injective.
\end{lem}

\proof
The $k$-th exterior power $U(n)\to\Aut(\Lambda^k\,\CC^n)$ 
of the canonical representation $U(n)\to\Aut(\CC^n)$ defines the element of $K^1(U(n))$; 
denote this element by $\beta_k$.
The ring $K^*(U(n))$ is the exterior algebra over $\Z$ 
generated by $\beta_1,\ldots,\beta_n$ \cite[Theorem 2.7.17]{Atiyah}.
Therefore, for every non-zero $\mu\in K^*(U(n))$ there is $\mu'\in K^*(U(n))$ 
such that $\mu\cdot\mu' = c_{\mu}b$, where $b=\beta_1\cdot\ldots\cdot\beta_n$ 
and $c_{\mu}$ is a non-zero integer. 
Thus the injectivity of $\pi^*$ is equivalent to the condition that 
$c\cdot\pi^* b\neq 0$ in $K^*(V\times T)$ for every integer $c\neq 0$.

By the K\"unneth formula \cite[Theorem 2.7.15]{Atiyah}, 
$K^*(T)$ is the exterior algebra over $\Z$ 
generated by the elements $\alpha_1, \ldots, \alpha_n \in K^1(T)$.
Applying the K\"unneth formula one more time, we obtain 
$K^*(V\times T) = K^*(V)\otimes K^*(T)$.
The group $K^*(T)$ is free Abelian and $K^*(V)$ is torsion-free,
so $K^*(V)\otimes K^*(T)$ is also torsion-free. 
Hence we should only prove that $\pi^* b\neq 0$.
Let us compute $\pi^* b$.
\begin{equation}\label{eq:rok}
\pi^*\beta_k 
  = \sum_{\substack{ I\subset\set{1,\ldots,n} \\ |I|=k} } \br{ \sum_{i\in I}\alpha_i \cdot \prod_{j\in I}l_j }
	= \sum_{i=1}^n{\alpha_i l_i} \sum_{\substack{I\subset\set{1,\ldots,n}\setminus\set{i} \\ |I|=k-1}} \prod_{j\in I}l_j
	= \sum_{i=1}^n{\alpha_i l_i \brr{\Lambda^{k-1}E_i}}, 
\end{equation}
where we denoted $E_i = \bigoplus_{j\neq i}L_j$.
Since $[L_i\oplus E_i] = n$, we have 
$\brr{\Lambda^{k}E_i} + l_i\brr{\Lambda^{k-1}E_i} = \brr{\Lambda^{k}(L_i\oplus E_i)} = \binom{n}{k}$,
where $\binom{n}{k}$ are the binomial coefficients.
Induction by $k$ gives $\brr{\Lambda^{k}E_i} = q_k(l_i)$,
where the polynomials $q_k\in\Z[x]$ are defined by the formula 
	$q_k(x) = \sum_{j=0}^k (-1)^j \binom{n}{k-j}x^j$.
Substituting this to \eqref{eq:rok}, we get $\pi^*\beta_k = \sum_{i=1}^n{\alpha_i l_i q_{k-1}(l_i)}$.
Taking the product of these identities for $k$ running from $1$ to $n$ 
and using the identity $\prod l_i = 1$, we obtain
\begin{equation}\label{eq:pib}
  \pi^*b = \prod_{k=1}^n\pi^*\beta_k = Q(l_1,\ldots,l_n)\cdot\alpha_1\cdot\ldots\cdot\alpha_n,
\end{equation}
where $Q\in\Z[x_1,\ldots,x_n]$ is the determinant of the matrix $(q_{k-1}(x_i))_{i,k=1..n}$.
Since $(-1)^k q_k(x)$ is a unital polynomial of degree $k$, 
the polynomial $Q$ is equal up to sign to the Vandermonde determinant 
$d_n(x_1,\ldots,x_n) = \det(x_i^{k-1}) = \prod_{i>j}(x_i-x_j)$.

It will be more convenient for us to use $u_k = l_k-1$ as the generators of $K^0(V)$ instead of $l_k$. 
The ring homomorphism $\Z[x_1,\ldots, x_n] \to K^0(V)$ sending $x_i$ to $u_i$ is surjective;
its kernel is the ideal $J_n$ generated by the elementary symmetric polynomials 
$\sigma_k(x_1,\ldots, x_n)$, $k=1\ldots,n$ \cite[Proposition 2.7.13]{Atiyah}.
Obviously, $d_n(l_1,\ldots,l_n) = \prod_{i>j}(l_i-l_j) = \prod_{i>j}(u_i-u_j) = d_n(u_1,\ldots,u_n)$.

Let us show that
\begin{equation}\label{eq:dn}
	d_n(x_1,\ldots,x_n) \equiv n!\prod_{k=1}^{n-1} x_{k+1}^{k} \;\mod J_n.
\end{equation}
Indeed, $d_2(x_1,x_2) = x_2-x_1 \equiv 2x_2 \;\mod J_2$. 
Let $n>2$ and suppose that 
\begin{equation}\label{eq:dn1}
d_{n-1}(x_1,\ldots,x_{n-1}) \equiv (n-1)!\prod_{k=1}^{n-2} x_{k+1}^{k} \;\mod J_{n-1}.
\end{equation}
Since $\sigma_k(x_1,\ldots,x_{n-1}) + x_n\sigma_{k-1}(x_1,\ldots,x_{n-1}) = \sigma_k(x_1,\ldots,x_{n}) \equiv 0 \,\mod J_n$,
induction by $k$ implies 
$\sigma_k(x_1,\ldots,x_{n-1}) \equiv (-1)^k x_n^k \;\mod J_n$
for all $k$. Hence
\begin{equation}\label{eq:nun}
	\prod_{1\leq j\leq n-1}(x_n-x_j) = \sum_{k=0}^{n-1}(-1)^k\sigma_k(x_1,\ldots,x_{n-1})x_n^{n-1-k} 
	  \equiv nx_n^{n-1} \;\mod J_n.
\end{equation}
The inverse image of the ideal $J_{n-1}$ under the projection 
$$\Z[x_1,\ldots,x_n]\to\Z[x_1,\ldots,x_n]/(x_n) = \Z[x_1,\ldots,x_{n-1}]$$
is the ideal generated by $x_n$ and $J_n$.
Taking into account induction assumption \eqref{eq:dn1}, we obtain
\begin{equation}\label{eq:app6}
\prod_{1\leq j<i\leq n-1}(x_i-x_j) \equiv (n-1)!\prod_{k=2}^{n-1} x_k^{k-1} +x_n f \;\mod J_{n}
\end{equation}
for some $f\in\Z[x_1,\ldots,x_n]$.
Multiplying \eqref{eq:app6} by \eqref{eq:nun}, we get 
\begin{equation}\label{eq:app7}
 \prod_{1\leq j<i\leq n}(x_i-x_j) \equiv n!\prod_{k=2}^{n} x_k^{k-1} + nf\cdot x_n^n \;\mod J_{n}. 
\end{equation}
Since $x_1,\ldots,x_n$ are roots of the polynomial $x^n - \sigma_1 x^{n-1} +\ldots + (-1)^n\sigma_n$, 
their $n$-th powers $x_i^n$ lie in $J_{n}$,
so $nf\cdot x_n^n \equiv 0 \;\mod J_{n}$,
and \eqref{eq:dn} follows from \eqref{eq:app7}.
Therefore, \eqref{eq:dn1} implies \eqref{eq:dn}, so \eqref{eq:dn} holds for all $n\geq 2$.

The quotient $\Z[x_1,\ldots,x_n]/J_n$ is a free Abelian group with the generators 
$\prod_{k=1}^{n-1} x_{k+1}^{j_k}$, $0\leq j_k\leq k$ \cite[Theorem 3.28]{Ka}.
The right-hand side of \eqref{eq:dn} coincides with one of these generators up to the factor $n!$,
so it does not vanish in $\Z[x_1,\ldots,x_n]/J_n$.
Equivalently, $d_n(u_1,\ldots,u_n)$ does not vanish in $K^0(V)$.
Taking into account that  $\alpha_1\cdot\ldots\cdot\alpha_n \neq 0$ in $K^*(T)$,
we finally obtain 
\begin{equation}\label{eq:pib2}
  \pi^*b = (-1)^{n(n-1)/2}\,n!\prod_{k=1}^{n-1} u_{k+1}^{k}\cdot\alpha_1\cdot\ldots\cdot\alpha_n 
	\neq 0 \text{ in } K^*(V\times T).
\end{equation}
This completes the proof of the lemma and of the proposition.
\endproof

\bigskip
\bigskip
\noindent
\textit{Department of Mathematics, Technion -- Israel Institute of Technology, Haifa, Israel} \\
\textit{marina.p@campus.technion.ac.il}


\begin{thebibliography}{9999}

\addtocontents{toc}{\vspace{8pt}}
\addcontentsline{toc}{section}{References}

\bibitem[A]{Atiyah}
M. F. Atiyah.
K-theory, Lecture notes by DW Anderson. 
Harvard, Benjamin, New York (Fall, 1964) (1967).

\bibitem[AB]{AB}
M. F. Atiyah and R. Bott.
The index problem for manifolds with boundary. 
Differential Analysis, Bombay Colloquium, Oxford Univ.Press, London (1964), 175--186.

\bibitem[AS1]{AS}
M. F. Atiyah and I. M. Singer.
Index theory for skew-adjoint Fredholm operators. 
Publications math\'ematiques de l'IH\'ES 37 (1969), no. 1, 5--26.

\bibitem[AS2]{AS71}
M. F. Atiyah and I. M. Singer.
The index of elliptic operators: IV. 
Annals of Mathematics 93 (1971), no. 1, 119--138.

\bibitem[APS]{APS-76}
M. F. Atiyah, V. K. Patodi, and I. M. Singer.
Spectral asymmetry and Riemannian geometry. III.
Mathematical Proceedings of the Cambridge Philosophical Society 79 (1976), no. 1, 71--99.

\bibitem[BLP]{BLP-04}
B. Booss-Bavnbek, M. Lesch, and J. Phillips. 
Unbounded Fredholm operators and spectral flow.
Canadian Journal of Mathematics 57 (2005), no. 2, 225--250; 
arXiv:math/0108014 [math.FA].

\bibitem[BdM]{Monvel}
L. Boutet de Monvel. 
Boundary problems for pseudo-differential operators. 
Acta mathematica 126 (1971), no. 1, 11--51. 

\bibitem[BJS]{BJS}
U. Bunke, M. Joachim, and S. Stolz. 
Classifying spaces and spectra representing the K-theory of a graded C*-algebra. 
High-dimensional manifold topology (2003), 80-102.

\bibitem[CL]{CL}
H. O. Cordes and J. P. Labrousse.
The invariance of the index in the metric space of closed operators.
Journal of Mathematics and Mechanics 12 (1963), no. 5, 693--719.

\bibitem[GL]{GL}
A. Gorokhovsky and M. Lesch.
On the spectral flow for Dirac operators with local boundary conditions.
International Mathematics Research Notices 17 (2015), 8036--8051;
arXiv:1310.0210 [math.AP].

\bibitem[J]{Jo}
M. Joachim.
Unbounded Fredholm operators and K-theory.
High-dimensional Manifold Topology, World Sci. Publ., River Edge, NJ (2003), 177--199.

\bibitem[Kar]{Ka}
M. Karoubi. K-theory. An introduction. Springer-Verlag (1978).

\bibitem[Kat]{Kato}
T. Kato. 
Perturbation Theory for Linear Operators. 
A Series of Comprehensive Studies in Mathematics 132 (1980).

\bibitem[KN]{KN}
M. I. Katsnelson and V. E. Nazaikinskii.
The Aharonov-Bohm effect for massless Dirac fermions and the spectral flow 
of Dirac type operators with classical boundary conditions.
Theoretical and Mathematical Physics 172 (2012), no. 3, 1263--1277;
arXiv:1204.2276 [math.AP].

\bibitem[Ku]{Kui}
N. H. Kuiper. 
The homotopy type of the unitary group of Hilbert space. 
Topology 3 (1965), no. 1, 19--30.

\bibitem[MSS]{MSS}
S. Melo, E. Schrohe, and T. Schick. 
Families index for Boutet de Monvel opera\-tors. 
M{\"u}nster Journal of Mathematics 6 (2013), no. 2, 343--364;
arXiv:1203.0482 [math.KT].

\bibitem[MP1]{MP1}
R.B. Melrose and P. Piazza.
Families of Dirac operators, boundaries and the b-calculus. 
Journal of Differential Geometry 46 (1997), no. 1, 99--180.

\bibitem[MP2]{MP2}
R.B. Melrose and P. Piazza.
An index theorem for families of Dirac operators on odd-dimensional manifolds with boundary. 
Journal of Differential Geometry 46 (1997), no.2, 287--334.

\bibitem[Pa]{Pal}
R. S. Palais. 
On the homotopy type of certain groups of operators. 
Topology 3 (1965), no. 3, 271--279.

\bibitem[P1]{Pr11}
M. Prokhorova.
The spectral flow for Dirac operators on compact planar domains with local boundary conditions.
Communications in Mathematical Physics 322 (2013), no.2, 385--414;
arXiv:1108.0806 [math-ph].

\bibitem[P2]{Pr15}
M. Prokhorova.
Index theorem for self-adjoint elliptic operators on a compact surface.
Abstracts of V School-Conference on algebraic geometry and complex analysis (August 2015, Koryazhma, Russia),
89--90.

\bibitem[P3]{Pr17}
M. Prokhorova.
Self-adjoint local boundary problems on compact surfaces. I. Spectral flow.
Journal of Geometric Analysis (2019), 
https://doi.org/10.1007/s12220-019-00313-0, 46 pp.; 
arXiv:1703.06105 [math.AP].

\bibitem[P4]{Pr19}
M. Prokhorova. Family index for unbounded operators. In preparation.

\bibitem[RS]{RS}
J. Robbin and D. Salamon.
The spectral flow and the Maslov index.
Bulletin of the London Mathematical Society 27 (1995), no. 1, 1--33.

\bibitem[Yu]{Yu}
Jianqing Yu.
Higher spectral flow for Dirac operators with local boundary conditions. 
International Journal of Mathematics 27 (2016), no. 8, 171--190;
arXiv:1410.5988 [math.DG]

\end{thebibliography}
\end{document}